\documentclass[final,twocolumn]{IEEEtran}

\usepackage[margin=2cm]{geometry}

\usepackage{cite}
\usepackage{url}
\usepackage{amssymb,amsmath}
\usepackage{array}
\usepackage{graphicx}
\usepackage{subfigure}

\begin{document}
%
\title{Distributed Soft Coding with a Soft Input Soft Output (SISO)
Relay Encoder in Parallel Relay Channels}

\author{Yonghui~Li,~\IEEEmembership{Senior~Member,~IEEE,} Md. Shahriar Rahman, Soon~Xin~Ng,~\IEEEmembership{Senior~Member,~IEEE,} and
Branka~Vucetic,~\IEEEmembership{Fellow,~IEEE}
\thanks{Copyright (c) 2010 IEEE. Personal use of this material is permitted. However, permission to use this material for any other purposes must be obtained from the IEEE by sending a request to pubs-permissions@ieee.org.}
\thanks{This work was supported by the Australian Research Council (ARC)
under Grants DP120100190, FT120100487, LP0991663, DP0877090.}
\thanks{Yonghui Li, Md. Shahriar Rahman and Branka Vucetic are with School of Electrical and Information Engineering,
        University of Sydney, Sydney, NSW, 2006, Australia. Tel: 61 2 9351 2236,
        Fax: 61 2 9351 3847. Email: \{yonghui.li,m.rahman,branka.vucetic\}@sydney.edu.au.}
\thanks{Soon Xin Ng is with the School of Electronics and
Computer Science, University of Southampton, SO17 1BJ, United
Kingdom. Email: sxn@ecs.soton.ac.uk.}}

\maketitle

\begin{abstract}
In this paper, we propose a new distributed coding structure with a
soft input soft output (SISO) relay encoder for error-prone parallel
relay channels. We refer to it as the distributed soft coding
(DISC). In the proposed scheme, each relay first uses the received
noisy signals to calculate the soft bit estimate (SBE) of the source
symbols. A simple SISO encoder is developed to encode the SBEs of
source symbols based on a constituent code generator matrix. The
SISO encoder outputs at different relays are then forwarded to the
destination and form a distributed codeword. The performance of the
proposed scheme is analyzed. It is shown that its performance is
determined by the generator sequence weight (GSW) of the relay
constituent codes, where the GSW of a constituent code is defined as
the number of ones in its generator sequence. A new coding design
criterion for optimally assigning the constituent codes to all the
relays is proposed based on the analysis. Results show that the
proposed DISC can effectively circumvent the error propagation due
to the decoding errors in the conventional detect and forward (DF)
with relay re-encoding and bring considerable coding gains, compared
to the conventional soft information relaying.
\end{abstract}

\begin{keywords}
Cooperative communications; Decode and Forward; Distributed Coding;
Relay Networks, Soft Coding.
\end{keywords}

\setcounter{page}{1}

\section{Introduction}
In wireless networks, the transmitted signal is overheard by all
nodes in the vicinity of the transmitter. Similarly, a receiver can
hear transmissions from multiple neighbouring nodes. This broadcast
nature of wireless networks provides unique opportunities for
collaborative and distributed signal processing techniques. Nodes
other than the intended destination can listen to a signal at no
additional transmission cost and it is globally efficient for these
nodes to forward the information to the destination. This process of
transmitting data from source to destination via one or more nodes
is referred to as relaying, which has been shown to yield spatial
diversity and great power savings [1-2].

Two most frequently used relaying protocols in relay networks are
amplify and forward (AF) and detect and forward (DF), which is also
referred to as the decode and forward for the coded systems [1-3,
37]. Recently, some variations of AF and DF protocols have been
proposed to further improve the performance of relayed transmission,
such as selective DF [1], soft information relaying (SIR) and
adaptive relaying protocol (ARP) [9, 43, 45]. Among them, the SIR
scheme has recently attracted a lot of interest due to its superior
performance [4-8, 23-28, 33]. There are generally various ways to
represent the soft information. Two commonly used SIR schemes are
the SIR scheme based on soft bit estimate (SIR-SBE) and the SIR
based on log-likelihood ratio (SIR-LLR).

The SIR-SBE schemes and soft encoding were first proposed in [4, 8],
where a soft encoding method was developed to calculate the SBE of
coded symbols. The SIR-LLR was first discussed in [5], where the
relay calculates and forwards the LLR of received symbols. It was
shown in [6, 31] that the SIR-SBE is an optimal relaying protocol in
terms of minimizing the mean square error (MSE) or maximizing the
overall destination SNR. The Gaussian distribution has been used to
approximate the probability distribution function (PDF) of both SBE
and LLR. However, it was later shown in [7, 29, 34] that the SBE of
an encoder output does not follow the exact Gaussian distributions,
which is particularly true for recursive encoders. To further
improve the performance of SIR scheme, [29] proposed an accurate
error model of SBE by dividing the error pattern in SBE into hard
and soft errors and calculating them separately. It is shown that
such error model is more accurate than the Gaussian distribution and
brings considerable performance improvement. Soft fading was
proposed in [36] as an alternative way to model the soft-errors
introduced in SIR processing at the relay via fading coefficients. A
mutual information based SIR scheme was developed in [33], where the
relay forwarding function consists of the hard decisions of the
symbol estimates and a reliability measure. The reliability measure
was determined by the symbol-wise mutual information computed from
the absolute value of the LLR at the relay. In [24], a novel
approach was proposed for analyzing the performance of SIR scheme by
calculating the mutual information loss due to the use of a soft
channel encoder at the relay. Soft information estimate and forward
scheme was further extended to higher order such as MQAM modulations
in [28] and to two way relay networks in [26]. In general, all these
schemes take advantage of soft information processing at the relay
to improve system performance.

It is well known that the main performance degradation in a DF
protocol is caused by the error propagation during the decoding and
relay encoding process when decoding errors occur at the relay.
Unlike DF protocol that makes a hard decision based on the
transmitted information symbols at the relay, SIR can effectively
alleviate error propagation by calculating and forwarding the
corresponding soft information. Forwarding soft information at the
relays provides additional information to the destination decoder to
make decisions, instead of making premature decisions at the relay
decoder. Thus SIR outperforms both AF and DF protocols and can
achieve a full diversity order in fading channels [33].

Recently, some distributed coding schemes have been proposed to
exploit the spatial diversity and the distributed coding gains in
wireless relay networks [47]. By applying the space time coding
principle, distributed space time codes [10-13, 49] have been
proposed for wireless relay networks. To further improve the system
performance, the distributed low density parity check (LDPC) codes
[14, 15, 18], distributed turbo coding (DTC) [16, 17, 42, 44, 46,
48], irregular distributed space-time code [49] and distributed
rateless coding [39-41] have been developed for a 2-hop single relay
network. However, these distributed coding schemes are based on the
DF relaying protocols and assume that either relays can always
decode correctly or relays will not forward when they cannot decode
correctly. Few of them have actually considered how to perform relay
encoding when there are decoding errors in the source to relay
links. In [29, 37], a few error models have been proposed for
distributed coding using DF protocol by taking into account the
decoding errors in the destination decoding. In [4, 7, 8, 23, 29,
31, 36, 38], distributed schemes with soft information encoding have
been developed for a relaying system containing only one relay when
imperfect decoding occurs at the relay. In these schemes, instead of
making decisions on the transmitted information symbols at the relay
as in other distributed coding schemes [14-18], the relay calculates
and forwards the corresponding soft estimates. The soft information
encoding methods proposed in these papers are basically a
probability inference method developed for the convolutional codes.
The encoding process is complicated and also it only considered one
relay. Very recently soft encoding was applied to distributed turbo
product and distributed LDPC codes in [35, 36]. It has been shown in
these papers that the soft information encoding can effectively
mitigate error propagation caused by the imperfect decoding and thus
improve the system performance.

To facilitate the hardware implementation, in [23] Winkelbauer and
Matz proposed a soft encoder structure implemented using shift
registers based on the boxplus operation. It has greatly reduced the
complexity of the soft encoder compared to the existing algorithms.
However, the boxplus representation in the soft encoder are still
quite complex and thus hard for performance analysis. Thus [23] only
considered a single relay system, and no performance analysis and
code design has been done. How to extend the soft encoding to
general multiple parallel relay network and design the optimal codes
for relay soft encoding remains an open problem.

Till now it is still unknown what is the optimal way to encode the
noisy estimates of the source symbols to circumvent error
propagation in the encoding process at the relay and simultaneously
provide significant distributed coding gains in general multiple
parallel relay network. It is still unclear if encoding these noisy
estimates can bring any further gains. A general framework for
designing and analyzing distributed coding in a general error-prone
relay channel is needed.

In this paper, we propose a simple distributed coding scheme with
soft input soft output (SISO) relay encoders for a two-hop parallel
relay network with one source, multiple parallel relays and one
destination. We refer to the proposed scheme as the distributed soft
coding (DISC). In DISC, different relays exploit a SISO encoder to
encode the received analog signals using different constituent
codes. The proposed SISO relay encoder has a very simple encoding
structure and can be implemented by using the structure of
conventional channel encoders in the complex number domain. Compared
to the SISO encoding scheme in [8], which is a probability inference
method and involves complicated calculations in encoding process,
the proposed SISO relay encoder can achieve the exactly the same
performance, but has a very simple encoding structure. It performs
the encoding of the noisy source symbol estimates at the relays.
Each relay first calculates the SBEs of the source symbols and the
SBEs are subsequently encoded by using the proposed SISO encoder,
whose outputs are then forwarded to the destination. At the
destination, the signals forwarded from various relays form a
distributed codeword.

\begin{figure*}
\centering
\includegraphics[width=0.7\textwidth]{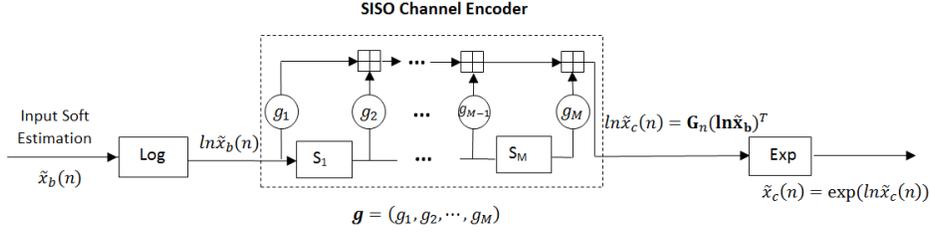}
\caption{SISO channel encoder, where the addition is done in the
complex field.}
\end{figure*}

The performance of the DISC is then analyzed. It is shown that its
performance is determined by the generator sequence weights (GSWs)
of the relay constituent codes, where the GSW of a constituent code
is defined as the number of ones in its generator sequence. To
optimize the BER performance, one should make the GSW of each
constituent code as large as possible by increasing the memory
length of the relay encoder. This is the same as for the
conventional convolutional codes. However, the GSWs cannot be chosen
arbitrarily, as the code may become catastrophic for certain
combinations of GSW values. For most of conventional noncatastrophic
convolutional codes, GSWs are different for different constituent
codes. Given K GSWs of K constituent codes, we need to determine
what is the optimal way of pairing off the K constituent codes with
K relays. A coding design criterion is proposed based on the BER
performance analysis of the DISC. It is shown that for the optimal
pairing one should match the GSW value of a constituent code to the
corresponding input SNRs of a relay and assign the constituent code
with a large GSW to the relay with a large input SNR. Simulation
results show that the proposed DISC with optimal pairing is superior
to the DISC with the un-ordered pairing. It is also shown that the
DISC can effectively overcome error propagation in the encoding
process at the relays and thus significantly outperforms the
conventional detect and forward (DF) schemes with relay re-encoding.
Furthermore, unlike the DF scheme, where the error performance
degrades as the number of encoder states increases, the performance
of the proposed DISC greatly improves as the number of encoder
states increases thus bringing significant distributed coding gains
compared to the conventional soft information relaying (SIR) without
relay encoding.

The rest of the paper is organized as follows. In Section II, we
describe the proposed DISC scheme. Its performance is analyzed in
Section III. Section IV provides the simulation results. Conclusions
are drawn in Section V.

\textbf{Notations}: In this paper, we use a lowercase and capital
bold face letter to denote a vector and a matrix, respectively. The
modulated signal of a symbol $b(n)$ is denoted as $x_b(n)$  and the
soft estimate of $b(n)$, given its a posteriori probability, is
denoted as $\tilde{x}_b(n)$.

\section{Distributed Soft Coding with a SISO Relay Encoder}

In this section, we first briefly describe a SISO encoder using a
single constituent convolutional code (CC) of rate 1. We then apply
it to a general two-hop relay network using multiple constituent
codes.

\subsection{SISO Convolutional Encoder}

Let us consider a rate-1 CC with the generator sequence of
$\mathbf{g}=(g_1,g_2,\ldots,g_M)$ and the generator matrix
$\mathbf{G}$. Let us denote by ${\bf{c}} = (c(1), \cdots ,c(N))$ the
encoder output codeword of information bits $\mathbf{b}= (b(1),
\cdots ,b(N))$, generated by the generator matrix $\mathbf{G}$,
where $c(n)$ is the $n$-th code symbol of $\mathbf{b}$. Let
${{\bf{x}}_c} = \left( {{x_c}(1),{x_c}(2), \cdots ,{x_c}(N)}
\right)$ represent the modulated sequence of $\mathbf{c}$.

Let $\mathbf{r}$ be a vector of analog signals, carrying information
bits $\mathbf{b}$. Let us denote the a posteriori probability (APP)
vector of $\mathbf{b}$, given $\mathbf{r}$, by
\begin{eqnarray}
{\bf{P}}{{\bf{r}}_{\bf{r}}} = \left\{ {\left( {{{Pr }_0}(1),{{Pr
}_1}(1)} \right), \cdots ,\left( {{{Pr }_0}(N),{{Pr }_1}(N)}
\right)} \right\}, \label{eq4}
\end{eqnarray}
where $Pr_0(n)$ and $Pr_1(n)$ are the APPs of  $b(n)=0$ and
$b(n)=1$, given $\mathbf{r}$ at time $n$. Given the APPs of
$\mathbf{b}$, the soft bit estimate (SBE) of $b(n)$, denoted by
${\tilde x_b}(n)$, representing the expected value of ${x_b}(n)$,
can be calculated as
\begin{eqnarray}
{\tilde x_b}(n) = E({x_b}(n)|{\bf{r}}) = {Pr_0}(n) - {Pr_1}(n).
\end{eqnarray}

As shown in [6, 8, 21], the SBE of symbol ${x_b}(n)$, denoted by
${\tilde x_b}(n)$, can be represented by the following model
\begin{eqnarray}
{\tilde x_b}(n) = {x_b}(n)\left( {1 - {{\tilde w}}(n)} \right) =
\alpha {x_b}(n) + {w_{in}}(n), \label{eq15}
\end{eqnarray}
where ${w_{in}}(n) =  - {x_b}(n)\left( {{{\tilde w}}(n) - {\mu
_{\tilde w}}} \right)$ is the noise term in the SBE with a zero mean
and variance of $\sigma _{in}^2$, ${\tilde w}(n) > 0$ is an
equivalent noise independent of ${x_b}(n)$ with a mean
\begin{eqnarray}
{\mu_{\tilde w}}=\frac{1}{N}\sum_{n=1}^N{{\tilde
w}(n)}=\frac{1}{N}\sum_{n=1}^N{(1-{x_b}(n){\tilde x_b}(n))},
\nonumber
\end{eqnarray}
and variance of
\begin{eqnarray}
\sigma _{\tilde w}^2=\frac{1}{N}\sum_{n=1}^N{((1-{x_b}(n){\tilde
x_b}(n)-\mu_{\tilde w}))^2},
\end{eqnarray}
and $\alpha = 1 - {\mu _{\tilde w}}$. It can be easily verified that
${x_b}(n)$ and ${w_{in}}(n)$ are independent.

%

Given $\mathbf{r}$, or equivalently, given the SBEs of information
bits $\mathbf{b}$, now let us look at how to calculate the SBEs of
the codeword $c(n)$ of $\mathbf{b}$, denoted by ${\tilde x_c}(n)$.


%

To calculate the SBEs of the codeword $\mathbf{c}$, let us first use
the probability inference method in [8] to calculate the
probabilities of $\mathbf{c}$, denoted by $Pr \left(
{{x_c}(n)|{{\bf{Pr_r}}}} \right)$, as follows
\begin{eqnarray}
Pr \left( {{x_c}(n) = q|{{\bf{Pr_r}}}} \right) = \sum\limits_{m \in
U({x_c}(n) = q)} {{{Pr }_{{b_c}}}\left( n \right)Pr \left(
{{S_m}(n)} \right)} \label{eq5}
\end{eqnarray}
\begin{eqnarray}
Pr \left( {{S_m}(n + 1)} \right) &=& \sum\limits_{m'} {Pr \left(
{{S_m}(n+1)|{S_{m'}}(n)} \right)Pr \left( {{S_{m'}}(n)} \right)}
\nonumber \\ &=& \sum\limits_{m'} {{{Pr }_{{b_n}(m,m')}}\left( n
\right)Pr \left( {{S_{m'}}(n)} \right)} \label{eq6}
\end{eqnarray}
where $S_m(n)$ represents the encoder states of $m$ at time $n$,
$U({x_c}(n) = q)$ is the set of branches, whose encoder output is
equal to $q$, $b_c$  is the corresponding input information bit,
${Pr _{{b_c}}}\left( n \right)$ is the probability of  $b_c$ at time
$n$, ${b_n}(m,m')$ represents the input information bit resulting in
the transition from state $m'$ at time $n$ to $m$ at time $n+1$, and
${Pr _{{b_n}(m,m')}}\left( n \right)$ is the probability of
information bit ${b_n}(m,m')$ at time $n$.

To gain more insights into the soft encoding process, let us first
consider a simple rate-1 CC with the generator sequence of
$\mathbf{g}=(1,~1,~1)$ and apply the above equations alternatively
to this code, we can easily find that the SBE of encoder input and
output for this code has the following simple relationship
\begin{eqnarray}
{\tilde x_c}(n)={\tilde x_b}(n){\tilde x_b}(n - 1){\tilde x_b}(n -
2) \label{eq8a}
\end{eqnarray}

Let us define ${\bf{ln}}{{\bf{\tilde x}}_b} = \left( {\ln {{\tilde
x}_b}(1), \cdots ,\ln {{\tilde x}_b}(N)} \right)$ and
${\bf{ln}}{{\bf{\tilde x}}_c} = \left( {\ln {{\tilde x}_c}(1),
\cdots ,\ln {{\tilde x}_c}(N)} \right)$, where for a complex number
$x = r{e^{j\theta }}$,  $\ln x = \ln |r| + j\theta$. Therefore, for
a negative real number $x<0$, $\ln x = \ln |x| + j\pi$. Then Eq.
(\ref{eq8a}) can be further written as
\begin{eqnarray}
\ln {\tilde x_c}(n) &=& \ln {\tilde x_b}(n) + \ln {\tilde x_b}(n -
1) + \ln {\tilde x_b}(n - 2) \nonumber \\ &=&{{\bf{G}}_n}\left(
{{\bf{ln}}{{{\bf{\tilde x}}}_b}} \right)^T \label{eq9}
\end{eqnarray}
\begin{eqnarray}
{\bf{ln}}{{\bf{\tilde x}}_c}=\mathbf{G} {\left(
{{\bf{ln}}{{{\bf{\tilde x}}}_b}} \right)^T} \label{eq10}
\end{eqnarray}
where ${{\bf{G}}_n}$ is the $n$-th row of $\mathbf{G}$.

(\ref{eq9}) and (\ref{eq10}) are derived for an example constituent
convolutional code (CC), but they can be applied to the general
non-recursive convolutional codes, as shown in the following
theorem.

\textbf{Theorem 1} - {\textbf{{SISO Encoder}}}: Let ${\tilde
x_b}\left(n \right)$ represent the input soft bit estimate (SBE) to
a channel encoder. Then given a generator matrix ${\bf{G}} = {\left(
{{\bf{G}^T_1},{\bf{G}^T_2}, \cdots ,{\bf{G}^T_N}} \right)^T}$ of a
rate-1 linear constituent code, where ${\bf{G}}_n$ is the $n$-th row
of \textbf{G} and the $ij$-th element of $\mathbf{G}$ is denoted by
${G_{ij}}$, ${G_{ij}} \in \left\{ {0,1} \right\}$, the logarithm of
the SISO encoder outputs for the soft inputs ${\bf{ln}}{{\bf{\tilde
x}}_b}$, denoted by ${\bf{ln}}{{\bf{\tilde x}}_c}$, can be
calculated as
\begin{eqnarray}
{\bf{ln}}{{\bf{\tilde x}}_c} = {\bf{G}}{\left(
{{\bf{ln}}{{{\bf{\tilde x}}}_b}} \right)^T}. \label{eq11}
\end{eqnarray}

The corresponding soft encoder outputs are given by
%
\begin{eqnarray}
{\tilde x_c}(n) &=& \exp \left( {\ln {{\tilde x}_c}(n)} \right) =
\exp \left( {{{\bf{G}}_n}{{\left( {{\bf{ln}}{{{\bf{\tilde x}}}_b}}
\right)}^T}} \right) \nonumber \\ &=& \exp \left( {\sum\limits_{i =
1}^N {{G_{ni}}\ln {{\tilde x}_b}(i)} } \right) = \prod\limits_{i \in
\{ {U_n}\} }^{} {{{\tilde x}_b}(i)} \label{eq13}
\end{eqnarray}
where ${U_n} = \{G_{nj} = 1,j = 1,...,N\}$, $n = 1,...,N$, is the
set of non-zero coefficients in ${\bf{G}}_n=({G_{n1}},{G_{n2}},
\cdots ,{G_{nN}})$.

 Proof: See Appendix A.

Theorem 1 formulates a simple SISO encoder structure, which is shown
in Fig. 1. As shown in the figure, the SISO encoder can be
implemented by using a convolutional encoder structure, but the
difference is that the addition operation in the SISO convolutional
encoder is done in a complex field, not in binary field. We can see
from the figure that the SISO channel encoder can be implemented by
adding a logarithm and an exponential module at the front and the
back of the conventional encoder, respectively.

\subsection{Distributed Soft Coding with a SISO Relay Encoder}

In this subsection, we introduce a distributed soft coding (DISC)
scheme for a general 2-hop parallel relay network based on the
previously described SISO encoder. For simplicity, in the following
sections, we consider a non-recursive convolutional code (NRCC) as
the constituent code in the SISO encoder at each relay. As shown in
Fig. 2, the network consists of one source, $K$ relays and one
destination. We assume that there is no direct link between the
source and destination as it is too weak compared to the relay links
and is thus ignored. For simplicity, we assume that no channel
coding is performed at the source node. Since we consider an uncoded
system, we will refer to DF as the detect and forward in the rest of
this paper. The proposed scheme can also be applied to the coded
system. If a channel code is applied at the source, relays will
perform the same soft encoding process as for an uncoded system, but
a concatenated code will be formed at the destination.

\begin{figure}
\centering
\includegraphics[width=0.4\textwidth]{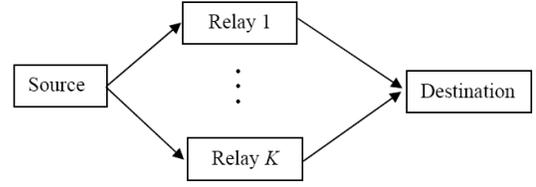}
\caption{A general two-hop parallel relay network.}
\end{figure}

Let $\mathbf{b}=(b(1),\cdots,b(n),\cdots,b(N))$ be a binary
information sequence of length $N$, generated by the source node. $\mathbf{b}$ is first modulated into a signal sequence
$\mathbf{x_b}=(x_b(1),\cdots,x_b(n),\cdots,x_b(N))$ and then
transmitted, where $x_b(n)$ is a modulated signal of $b(n)$. For
simplicity, in this paper, we consider the BPSK modulation and
assume that the symbol 0 and 1 are modulated into 1 and -1,
respectively. Similar to [1, 2, 8, 9], we assume that the source and
relays transmit signals over orthogonal channels. We will
concentrate on a time division scheme, for which each node transmits
in a separate time slot. The source first broadcasts the signals
$\mathbf{x_b}$ to all $K$ parallel relays. Let ${{\bf{r}}_{s{r_k}}}
= \left( {r_{s{r_k}}^{}(1), \cdots ,r_{s{r_k}}^{}(N)} \right)$ be
the signals received at relay $k$, where
\begin{eqnarray}
r_{s{r_k}}(n) = \sqrt {{P_{s{r_k}}}} h_{s{r_k}}{x_b}(n) + \eta
_{{r_k}}(n), \label{eq1}
\end{eqnarray}
${P_{s{r_k}}} = {P_s}{L_{s{r_k}}}$ is the average power received at
relay $k$, $P_s$ is the source transmit power,  $L_{s{r_k}}$ and
$h_{s{r_k}}$ are the pathloss and channel gain between the source
and relay $k$, and $\eta _{{r_k}}(n)$ is a zero mean complex
Gaussian noise with variance of $\sigma _n^2$.

Let ${\bf{g}}^{(k)}$ and $\mathbf{G}^{(k)}$ represent the generator
sequence and generator matrix for relay $k$. $d_k$ denotes the
number of ones in ${\bf{g}}^{(k)}$ and is referred to as the row
degree of $\mathbf{G}^{(k)}$. Let ${{\bf{c}}_k} = ({c_k}(1), \cdots
,{c_k}(N))$ be the codeword of information bits \textbf{b} generated
by $\mathbf{G}^{(k)}$, where $c_k(n)$ is the $n$-th code symbol of
\textbf{b}. Let ${{\bf{x}}_{{c_k}}} = \left(
{{x_{{c_k}}}(1),{x_{{c_k}}}(2), \cdots ,{x_{{c_k}}}(N)} \right)$
represent the modulated sequence of $\mathbf{c}_k$.

\begin{figure}
\centering
\includegraphics[width=0.5\textwidth]{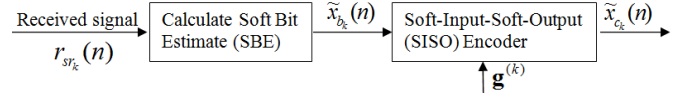}
\caption{The encoder structure at relay k in DISC.}
\end{figure}

Fig. 3 shows the proposed encoder structure at relay $k$. In the
proposed DISC scheme, each relay exploits a rate-1 SISO encoder to
encode the received analog signals based on its assigned constituent
code. It has been shown in [6] and [31] that the soft bit estimate
of source information symbol is the optimal input to the soft
encoder in terms of maximizing the SNR of the SISO encoder output.
Thus upon receiving signals, relay $k$ first calculates the SBE of
the source information symbol, denoted by ${\tilde x_{{b_k}}}(n)$.
The relay $k$ then applies the SISO encoder described in Theorem 1
to encode ${\tilde x_{{b_k}}}(n)$, based on the given generator
sequence ${\bf{g}}^{(k)}$, to generate the SBEs of its code sequence
${{\bf{c}}_k}$. Let $\tilde x_{{c_k}}(n)$  be the corresponding soft
encoder output at relay $k$ at time $n$. By substituting
(\ref{eq15}) into (\ref{eq13}), $\tilde x_{{c_k}}(n)$ can be
expressed as
\begin{eqnarray}
\tilde x_{{c_k}}^{}(n) &=& \prod\limits_{j \in \{ {U^{(k)}_n}\} }
{{{\tilde x}_{b_k}}(j)} \nonumber \\ &=& \prod\limits_{j \in \{
{U^{(k)}_n}\} }^{}
{\left( {\alpha_k {x_b}(j) + {w_{in,k}}(j)} \right)}  \nonumber \\
&=& {\alpha_k} ^{{d_k}} {\prod\limits_{j \in \{ {U^{(k)}_n}\}
}{x_b}(j)} + {w_{out,k}}(n) \\ \nonumber &=& {\alpha
_k}^{{d_k}}{x_{{c_k}}}(n) + {w_{out,k}}(n), \label{eq17}
\end{eqnarray}
where ${U^{(k)}_n} = \{G^{(k)}_{nj} = 1,j = 1,...,N\}$, $n =
1,...,N$, is the set of non-zero coefficients in
${\bf{G}}^{(k)}_n=({G^{(k)}_{n1}},{G^{(k)}_{n2}}, \cdots
,{G^{(k)}_{nN}})$; ${x_{c_k}}(n) = \prod\limits_{j \in \{
{U^{(k)}_n}\} }{{x_b}(j)}$  is the modulated signal of $n$-th code
bit $c_k(n)$ of $\mathbf{b}$; ${\alpha _k} = 1 - {\mu _{\tilde
w,k}}$; $\mu _{\tilde w,k}$ is the mean of equivalent noise ${\tilde
w}(n)$ in the SBE at relay $k$ shown in (\ref{eq15}), and
${w_{out,k}}(n)$ is the equivalent noise at the soft encoder output
with a  zero mean and variance of
\begin{eqnarray}
\sigma _{out,k}^2={\left( {\alpha _k^2 + \sigma _{in,k}^2}
\right)^{{d_k}}} - \alpha _k^{2{d_k}}, \label{eq21}
\end{eqnarray}
$\sigma _{in,k}^2$ is the variance of the equivalent noise
${w_{in}}(n)$ in the soft input to the SISO encoder of relay $k$, as
shown in (\ref{eq15}).

For simplicity, we assume that all relays transmit at the same power
$P_r$. Then the signals transmitted from relay $k$ can then be
written as
\begin{eqnarray}
x_{{r_k}}(n) = {\beta _k}\tilde x_{{c_k}}(n), \label{eq22}
\end{eqnarray}
where ${\beta _k}$ is a normalization factor, given by
\begin{eqnarray}
\beta _k= \sqrt {{P_r}/{P_{{x_k}}}}, \label{eq23}
\end{eqnarray}
where ${P_{{x_k}}} = E\left( {{{\left| {\tilde x_{{c_k}}(n)}
\right|}^2}} \right) = {\left( {\alpha _k^2 + \sigma _{in,k}^2}
\right)^{{d_k}}}$.

The destination received signal, transmitted from relay $k$, can be
written as
\begin{eqnarray}
y_{{r_k}d}(n) = \sqrt {{L_r}} h_{{r_k}d}x_{{r_k}}(n) +
w_{{r_k}d}(n) ~~~~~~~~~~~~~~~~~ \nonumber \\
= \sqrt {{L_r}} h_{{r_kd}}{\beta _k}\left( {\alpha
_k^{{d_k}}{{\tilde
x}_{{c_k}}}(n) + {w_{out,k}}(n)} \right) + w_{{r_k}d}(n) \nonumber \\
= \sqrt {{L_r}} h_{{r_k}d}{\beta _k}\alpha _k^{{d_k}}{\tilde
x_{{c_k}}}(n) + \tilde w_{{r_k}d}^{}(n),
~~~~~~~~~~~~~~~~~~\label{eq24}
\end{eqnarray}
where $L_r$ is the pathloss from the relay to the destination,
$\tilde w_{{r_k}d}(n) = \sqrt {{L_r}} h_{{r_k}d}{\beta
_k}{w_{out,k}}(n) + w_{{r_k}d}(n)$ is an equivalent noise with a
zero mean and variance of
\begin{eqnarray}
\sigma _{{r_k}d}^2 = \sigma _n^2 + L_r{\left| {h_{{r_k}d}}
\right|^2}\beta _k^2\left[ {{{\left( {\alpha _k^2 + \sigma
_{in,k}^2} \right)}^{{d_k}}} - \alpha _k^{2{d_k}}} \right].
\label{eq25}
\end{eqnarray}

In the above equation, we have used the fact that $E\left(
{{w_{in,i}}(n){w_{in,j}}(n)} \right) = 0$  for the BPSK
constellation [6] and thus $E\left( {{w_{out,i}}(n){w_{out,j}}(n)}
\right) = 0$.

As shown in \cite{7}\cite{34}, the noise term in the SBE
$\omega_{in}(n)$ does not follow the exact Gaussian distribution, so
does the equivalent noise of the soft encoder output
$\omega_{out}(n)$ and the overall destination noise
$\tilde{\omega}_{r_kd}(n)$. Fig. 4 plots the probability
distribution function (PDF) of overall destination noise for the
rate-1 4-state SISO encoder at various SNRs, which is also the
conditional PDF of $Pr(y_{{r_k}d}(n)|{x_{c_k}}(n))$. The PDF curves
are obtained by averaging over 1000 frames, each of which consists
of 130 symbols. It can be seen from the figure that the overall
destination noise can be roughly approximated by Gaussian
distributions with some approximation errors. To simplify the
decoding process and analysis, throughout the paper we assume that
the overall destination noise follows the Gaussian distribution.
Therefore, the conventional BCJR MAP decoding algorithm \cite{32}
can be directly applied to decode the destination received signals
by treating the overall received signals as a codeword of rate
$1/K$.

\begin{figure}
\centering \subfigure[$\bar \gamma _{s{r_1}}=\bar \gamma
_{s{r_2}}=\bar \gamma _{rd}=6dB$;]{
\includegraphics[width=0.48\textwidth]{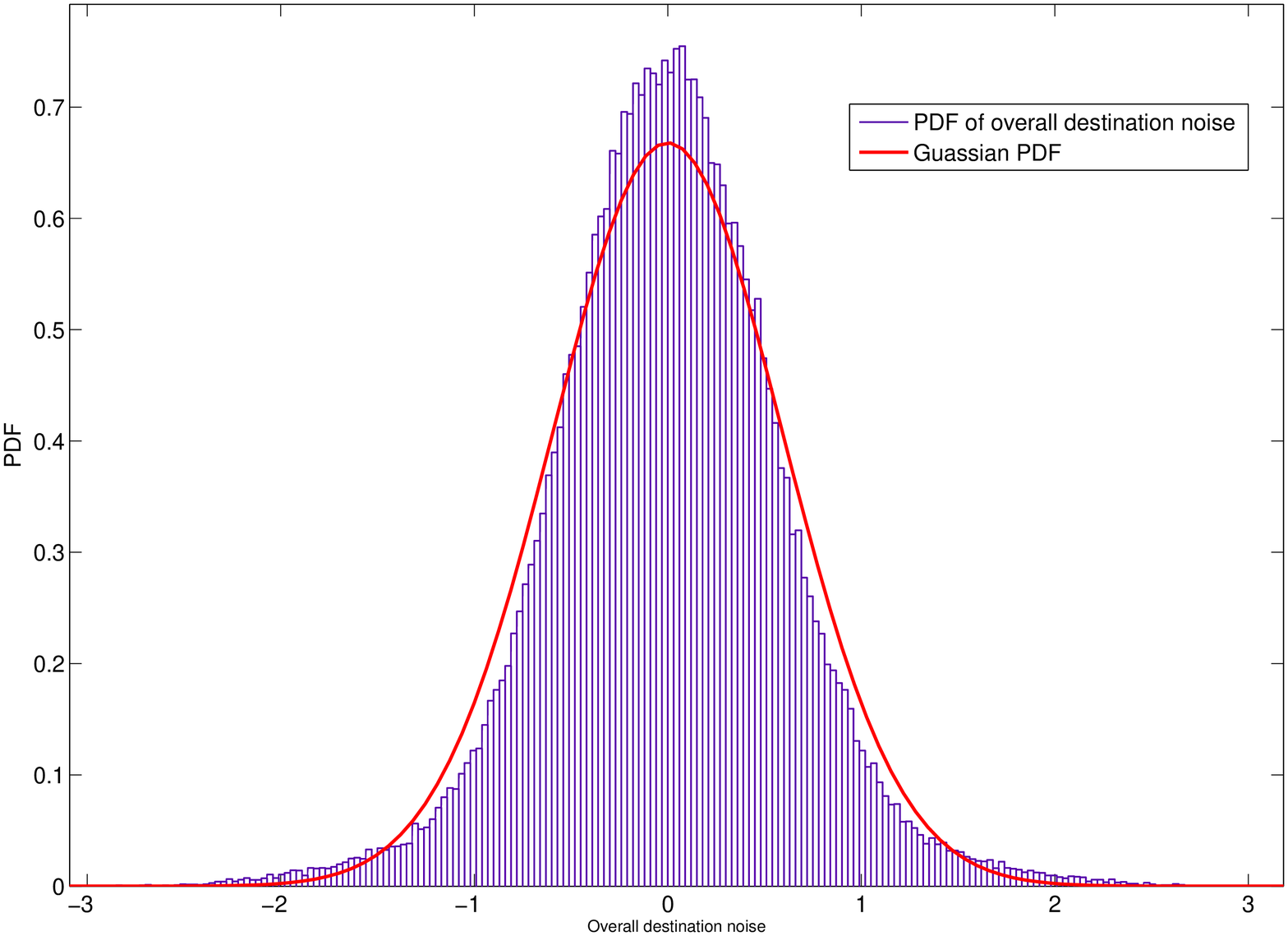}
\label{fig4_1}} \subfigure[$\bar \gamma _{s{r_1}}=\bar \gamma
_{s{r_2}}=\bar \gamma _{rd}=8dB$;]{
\includegraphics[width=0.48\textwidth]{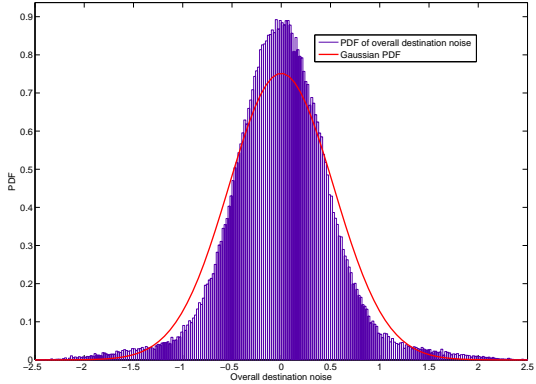}
\label{fig4_2}} \subfigure[$\bar \gamma _{s{r_1}}=\bar \gamma
_{s{r_2}}=\bar \gamma _{rd}=10dB$;]{
\includegraphics[width=0.48\textwidth]{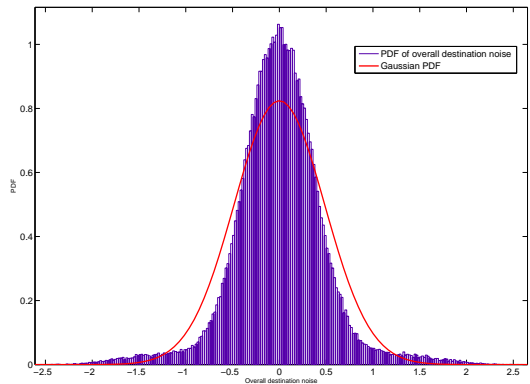} \label{fig4_3}}
\caption{PDF of overall destination noise for the 4-state SISO
encoder.}
\end{figure}

\section{Performance Analysis and DISC Design}

The SISO encoding at the relays can bring system some coding gains,
but on the other hand soft encoding of noisy SBE will enhance the
noise power at the destination, as shown in (\ref{eq17}). It is
unclear whether the coding gain can surpass the noise enhancement in
the DISC and if such encoding can bring any overall gain. Moreover,
given $K$ constituent codes, we need to determine what is the
optimal way of pairing off the $K$ constituent codes with $K$
relays.  In this section, we will give a quantitative analysis of
the DISC scheme performance based on which a coding design criterion
is then proposed.

By substituting (\ref{eq23}) into (\ref{eq25}), the variance of the
equivalent destination noise in Eq. (\ref{eq25}) can be expanded as
\begin{eqnarray}
\sigma _{{r_k}d}^2 = \sigma _n^2 + L_r^{}{\left| {h_{{r_k}d}^{}}
\right|^2}\beta _k^2\left[ {{{\left( {\alpha _k^2 + \sigma
_{in,k}^2} \right)}^{{d_k}}} - \alpha _k^{2{d_k}}} \right] \nonumber \\
= \sigma _n^2 + {P_r}{L_r}{\left| {h_{{r_k}d}^{}} \right|^2}\left[
{1 - {{\left( {1 + 1/{\gamma _{s{r_k},in}}} \right)}^{ - {d_k}}}}
\right], \label{eq26}
\end{eqnarray}
where ${\gamma _{s{r_k},in}} = \alpha _k^2/\sigma _{in,k}^2$ is the
input SNR of the SISO encoder at the $k$-th relay.

Let ${\gamma _{s{r_k}d}}$ represent the instantaneous destination
SNR, corresponding to the signals generated by the constituent
encoder at the $k$-th relay. Then from (\ref{eq24}) and
(\ref{eq26}), we get
\begin{eqnarray}
{\gamma _{s{r_k}d}} &=& \frac{{{L_r}{{\left| {h_{{r_k}d}{\beta
_k}\alpha _k^{{d_k}}} \right|}^2}}}{{\sigma _{{r_k}d}^2}} \nonumber
\\ &=& \frac{{{{{L_r}{P_r}{{\left| {h_{{r_k}d}^{}} \right|}^2}\alpha
_k^{2{d_k}}} \mathord{\left/ {\vphantom {{{L_r}{P_r}{{\left|
{h_{{r_k}d}} \right|}^2}\alpha _k^{2{d_k}}} {{{\left( {\alpha _k^2 +
\sigma _{in,k}^2} \right)}^{{d_k}}}}}} \right.
\kern-\nulldelimiterspace} {{{\left( {\alpha _k^2 + \sigma
_{in,k}^2} \right)}^{{d_k}}}}}}}{{\sigma _n^2 + {L_r}{P_r}{{\left|
{h_{{r_k}d}} \right|}^2}\left[ {1 - {{\left( {1 + 1/{\gamma
_{s{r_k},in}}} \right)}^{ - {d_k}}}} \right]}} \nonumber \\ &=&
\frac{{{\gamma _{{r_k}d}}}}{{\left[ {{{\left( {1 + 1/{\gamma
_{s{r_k},in}}} \right)}^{{d_k}}} - 1} \right]{\gamma _{r_kd}} +
\left( {1 + 1/{\gamma _{s{r_k},in}}} \right)^{d_k}}} \nonumber
\\ &\approx& \frac{{{\gamma _{{r_k}d}}{\gamma
_{s{r_k},in}}}}{{{d_k}{\gamma _{{r_k}d}} + {\gamma _{s{r_k},in}}}},
\label{eq27}
\end{eqnarray}
where  ${\gamma _{{r_k}d}} = {P_r}{L_r}{\left| {h_{{r_k}d}}
\right|^2}/\sigma _n^2$ is the instantaneous SNR in the link from
relay $k$ to the destination and the last equation is a high SNR
approximation.

Since calculating the exact BER is extremely complicated, we
consider an asymptotic performance at high SNR, which can give us
some insights into the system design. We assume that $\tilde
w_{{r_k}d}(n)$ can be approximated as a Gaussian random variable.
Then the instantaneous BER of the DISC scheme can be approximated at
high SNR as [21]
\begin{eqnarray}
{P_b} &\approx& {B_{{d_{free}}}}Q\left( {\sqrt {2\sum\limits_{k =
1}^K {{d_{\min ,k}}{\gamma _{s{r_k}d}}} } } \right) \nonumber
\\ &\approx&
0.5{B_{{d_{free}}}}\exp \left( { - \sum\limits_{k = 1}^K {{d_{\min
,k}}{\gamma _{s{r_k}d}}} } \right), \label{eq28}
\end{eqnarray}
where ${d_{\min ,k}}$  is the minimum Hamming weight (MHW) of a
nonzero codeword, which is also equal to the code minimum Hamming
distance (MHD), generated by the constituent encoder of relay $k$.

Let ${d_{\min }}$ denote the MHW of a nonzero codeword generated by
all $K$ constituent encoders. ${d_{\min }}$  and  ${d_{\min ,k}}$
can be obtained either by simulations or by deriving its bounds.
Theorem 2 presents a simple bound for ${d_{\min }}$  and  ${d_{\min
,k}}$.

 \textbf{Theorem 2}: Let us consider a non-recursive convolutional
code $C$, generated by $K$ constituent codes  ${\bf{g}}^{(1)},
{\bf{g}}^{(2)}, \cdots, {\bf{g}}^{(K)}$. Let ${d_{\min ,k}}$
represent the MHD of the code generated by the $k$-th constituent
code ${\bf{g}}^{(k)}$. Let denote by ${d_{\min }}$ the MHD of the
overall codeword generated by $K$ constituent codes. Also let
${w_{1,k}}$ be the Hamming weight of a codeword generated by the
$k$-th constituent encoder for the input sequence of $(1~0~0~0\cdots
0)$. Then we have the following simple bound for ${d_{\min }}$  and
${d_{\min ,k}}$,
\begin{eqnarray}
{d_{\min ,k}}\leq {w_{1,k}}=d_k, ~~{d_{\min }} \le \sum\limits_{k =
1}^K {{d_k}}, \label{eq29}
\end{eqnarray}
where $d_k$ is the row degree of the generator matrix for the $k$-th
constituent code, which is equal to the number of 1s in its
generator sequence ${\bf{g}}^{(k)}$. We refer to $d_k$  as the
generator sequence weight (GSW) of ${\bf{g}}^{(k)}$.

\begin{table*}
\centering \caption{Comparison of Exact MHD and MHD Bound in Theorem
2 for Rate 1/2, 1/3 and 1/4 codes, where the generator polynomial
are given in octal form.}
\includegraphics[width=1\textwidth]{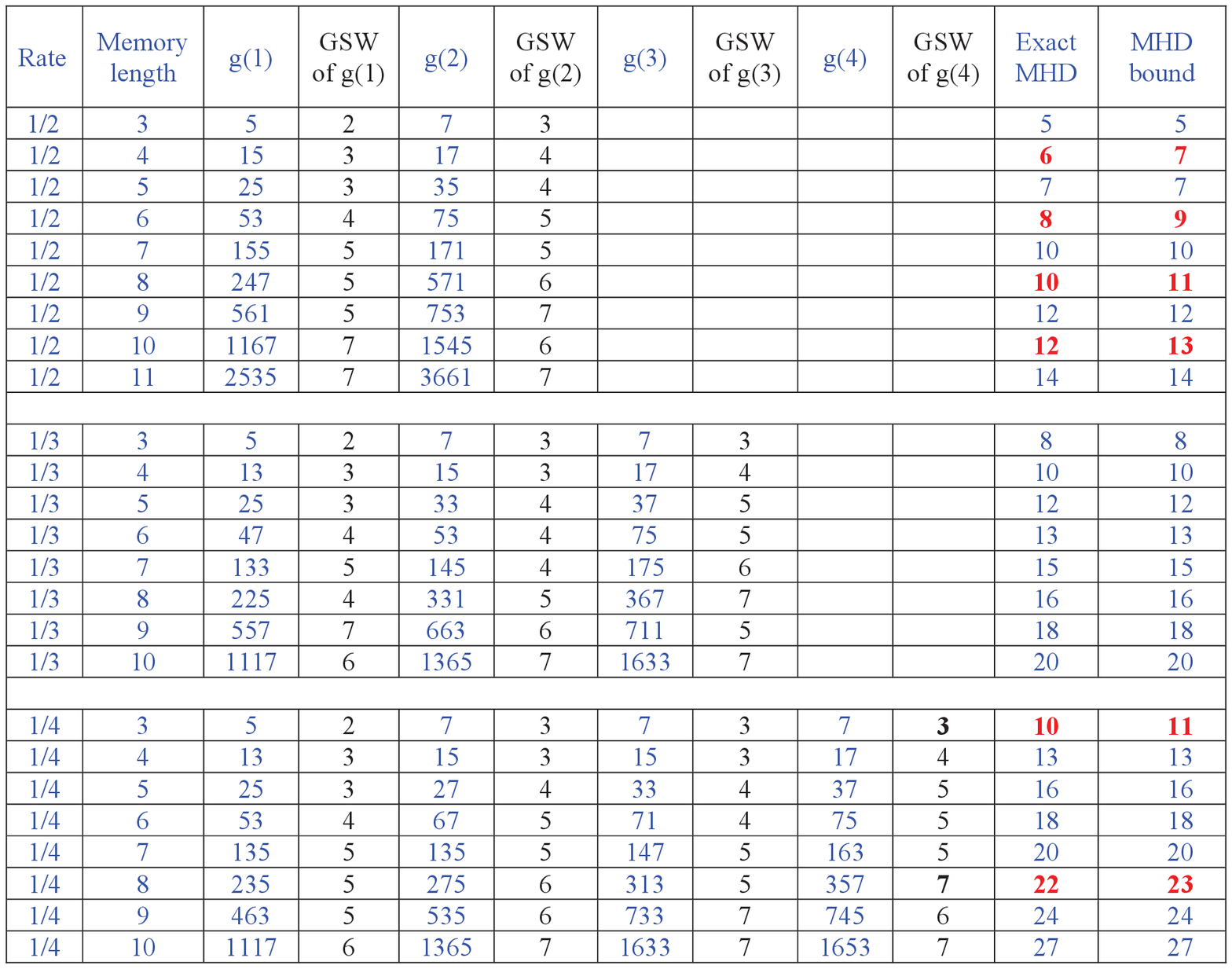}
\end{table*}

Proof of this bound is straightforward and we omit it here. It can
be derived based on the property of linear codes that the MHD of a
linear code is equal to the minimum Hamming weight of all non-zero
codewords. The bound in (\ref{eq29}) is an upper bound because it is
the Hamming weight of a specific codeword generated by a specific
input sequence $(1~0~0~0\cdots 0)$. Table 1 shows the exact MHDs and
the MHD bounds calculated in (\ref{eq29}) for rate 1/2, 1/3 and 1/4
codes with various memory lengths. The codes are obtained from [20].
We can see that the difference between the bound in Eq. (\ref{eq29})
and the exact MHD is at most 1 for the codes listed in the table and
for most of the codes the bound is equal to the exact MHD.

By using the MHD bound in Theorem 2, the instantaneous BER in Eq.
(\ref{eq28}) can be further approximated as
\begin{eqnarray}
{P_b} &\approx& 0.5{B_{{d_{free}}}}\exp \left( { - \sum\limits_{k =
1}^K {{d_{\min ,k}}{\gamma _{s{r_k}d}}} } \right) \nonumber
\\ &\approx&
0.5{B_{{d_{free}}}}\exp \left( { - \sum\limits_{k = 1}^K
{{d_k}{\gamma _{s{r_k}d}}} } \right) \nonumber \\ &\approx&
0.5{B_{{d_{free}}}}\exp \left( { - \sum\limits_{k = 1}^K
{\frac{{{d_k}{\gamma _{{r_k}d}}{\gamma _{s{r_k},in}}}}{{{d_k}{\gamma
_{{r_k}d}} + {\gamma _{s{r_k},in}}}}} } \right) \nonumber \\ &=&
0.5{B_{{d_{free}}}}\exp \left( { - \sum\limits_{k = 1}^K
{\frac{1}{{\gamma _{{r_k}d}^{ - 1}d_k^{ - 1} + \gamma _{s{r_k},in}^{
- 1}}}} } \right) \nonumber \\ &=& 0.5{B_{{d_{free}}}}\exp \left( {
- \sum\limits_{k = 1}^K {{\rho _k}\left( {{d_k},{\gamma
_{{r_k}d}},{\gamma _{s{r_k},in}}} \right)} } \right) ~~\label{eq30}
\end{eqnarray}
where
\begin{eqnarray}
{\rho _k}\left( {{d_k},{\gamma _{{r_k}d}},{\gamma _{sr,in,k}}}
\right) = \frac{{{\gamma _{{r_k}d}}{\gamma _{s{r_k},in}}}}{{{\gamma
_{s{r_k},in}}d_k^{ - 1} + {\gamma _{{r_k}d}}}}. \label{eq31}
\end{eqnarray}

If we substitute $d_{min,k}$=1 in (\ref{eq30}), we will obtain the
BER expression of the soft information relaying (SIR) scheme. As it
can be seen from (\ref{eq30}), the DISC scheme always outperforms
the SIR scheme. Since the SIR scheme can achieve the full diversity
order of $K$, the DISC scheme can also achieve the full diversity
order.

We can see from (\ref{eq30}-\ref{eq31}) that $P_b$ and ${\rho
_k}\left( {{d_k},{\gamma _{{r_k}d}},{\gamma _{s{r_k},in}}} \right)$
are a monotonic decreasing and increasing function of $d_k$,
respectively. To reduce the error rate $P_b$, one should make the
GSW $d_k$ as large as possible by increasing its memory length, as
with the conventional convolutional codes. However, $d_k$,
$k=1,\ldots,K$ cannot be chosen arbitrarily as the code may become
catastrophic for certain combinations of GSW values. The code
construction has to be non-catastrophic. Some examples of good
non-catastrophic codes are shown in Table 1 for various code rates.
We can see from these tables that for most of good codes, GSWs are
different for different $k$. Given $K$ GSWs of $K$ constituent codes
$(d_1,d_2,\ldots,d_K)$, we need to determine what is the optimal way
of pairing off the constituent codes with relays. That is, which
constituent code should be used in which relay?

To answer this question, let us first rearrange ${\gamma _{s{r_k},in}}$
in a decreasing order and denote the reordered SNR values as $\left(
{{\gamma _{s{r_{(1)}},in}},{\gamma _{s{r_{(2)}},in}}, \cdots
,{\gamma _{s{r_{(K)}},in}}} \right)$, where
\begin{eqnarray}
{\gamma _{s{r_{(1)}},in}} \ge {\gamma _{s{r_{(2)}},in}} \ge  \cdots
\ge {\gamma _{s{r_{(K)}},in}} \label{eq32}.
\end{eqnarray}

Similarly we also rearrange $(d_1,d_2,\ldots,d_K)$ in a decreasing
order as follows,
\begin{eqnarray}
{d_{(1)}} \ge {d_{(2)}} \ge  \cdots  \ge {d_{(K)}}. \label{eq33}
\end{eqnarray}

Then we have the following theorem regarding optimally pairing the
constituent codes with the relays for AWGN channels.

 \textbf{Theorem 3} - \textit{Design Criterion of DISC for AWGN Channels}: Let
us consider a parallel relay network consisting of K relays over
AWGN channels. We assume that all relays have the same transmission
power and experience the same path loss, that is, ${\gamma
_{{r_k}d}} = {\gamma _{rd}}$  for all $k=1,\ldots,K$. In the DISC,
each relay performs a SISO encoding, where the existing good
convolutional codes or other linear codes can be chosen as the relay
constituent codes. Assume that a good convolutional code generated
by $K$ constituent convolutional codes ${\bf{g}}^{(1)},
{\bf{g}}^{(2)}, \cdots, {\bf{g}}^{(K)}$,  has already been found.
Let us denote by $d_k$  the GSW of  ${\bf{g}}^{(k)}$. Then the
optimal code construction in AWGN channels is to assign the code
with the $k$-th largest GSW $d_{(k)}$ to the relay node with the
$k$-th largest input SNR ${\gamma _{s{r_{(k)}},in}}$. In this way,
the system achieves the optimal BER performance for given
constituent codes with generator sequences ${\bf{g}}^{(1)},
{\bf{g}}^{(2)}, \cdots, {\bf{g}}^{(K)}$.

         Proof: See Appendix B.

From the above theorem, we can see that in order to minimize the BER
$P_b$ for the given $K$ GSWs $(d_{(1)}, d_{(2)},\ldots,d_{(K)})$, we
should match the GSW values to the corresponding input SNRs and
assign the constituent code with a large GSW value to the relay with
a large SNR.

Let ${\bar \gamma _{s{r_k}}}$ be the average SNR in the link from
the source to relay $k$. Then by using the fact that ${\gamma
_{s{r_{(k)}},in}}$ is monotonically increasing function of ${\bar
\gamma _{s{r_k}}}$, we have the following Corollary.

   \textbf{Corollary 1}: The optimal code construction of DISC for AWGN
channels is to assign the code with the $k$-th largest GSW
${d_{(k)}}$ to the relay node with the $k$-th largest average SNR
${\bar \gamma _{s{r_k}}}$.

To show the gain of the DISC with the optimal pairing over a DISC
with un-ordered pairing and the soft information relaying (SIR), let
us consider a simple example.

\textbf{Example 2}: We consider a rate $1/2$ convolutional code of
memory length of 3. The generator sequences of two constituent codes
are ${\bf{g}}^{(1)}$=(101) and  ${\bf{g}}^{(2)}$=(111). Then we have
${d_{(1)}} = 3$  and  ${d_{(2)}} = 2$.

We consider an AWGN channel and assume that the average input SNR of
relay 1 is larger than that of relay 2. Then we have  ${\gamma
_{s{r_1},in}}>{\gamma _{s{r_2},in}}$.

Then according to Corollary 1 the optimal pairing strategy is to
assign the constituent code  ${\bf{g}}^{(2)}$=(111) with the maximum
GSW to relay 1 with the maximum SNR ${\gamma _{s{r_1},in}}$ and
${\bf{g}}^{(1)}$=(101) with the minimum GSW to relay 2 with the
minimum SNR ${\gamma _{s{r_2},in}}$.

To show the performance gain of the proposed scheme, we assume that
${\gamma _{s{r_1},in}} = {\gamma _{gap}}{\gamma _{rd}}$, where
${\gamma _{gap}} = {\gamma _{s{r_1},in}}/{\gamma _{rd}}$ and
${\gamma _{s{r_1},in}}/{\gamma _{s{r_2},in}} = {\alpha _0}$,
${\alpha _0}>1$. Then for an un-ordered pairing, we assume that
${\bf{g}}^{(1)}$ and ${\bf{g}}^{(2)}$ are assigned to relays 1 and
2, respectively. Then we have ${d_1} = 2$ and ${d_2} = 3$ and
\begin{eqnarray}
\rho _1^{un - order}\left( {d_1,{\gamma _{rd}},{\gamma
_{s{r_1},in}}} \right) &=& \frac{{\gamma _{gap}^{ - 1}{\gamma
_{s{r_1},in}}{\gamma _{s{r_1},in}}}}{{\left( {{\gamma
_{s{r_1},in}}d_1^{-1} + \gamma _{gap}^{ - 1}{\gamma _{s{r_1},in}}}
\right)}} \nonumber \\ &=& \frac{{{\gamma _{s{r_1},in}}}}{{\gamma
_{gap}/2 + 1}};
\end{eqnarray}
\begin{eqnarray}
\rho _2^{un - order}\left( {d_2,{\gamma _{rd}},{\gamma
_{s{r_2},in}}} \right) &=& \frac{{\gamma _{gap}^{ - 1}{\gamma
_{s{r_1},in}}{\gamma _{s{r_2},in}}}}{{\left( {{\gamma
_{s{r_2},in}}d_2^{- 1} + \gamma _{gap}^{ - 1}{\gamma _{s{r_1},in}}}
\right)}} \nonumber \\ &=& \frac{{{\gamma _{s{r_1},in}}}}{{1/3\gamma
_{gap} + {\alpha _0}}};
\end{eqnarray}
\begin{eqnarray}
\rho^{un - order} &=& \sum\limits_{k = 1}^2 {\rho_k^{un -
order}\left( {d_k^{},{\gamma _{rd}},{\gamma _{s{r_k},in}}} \right)}
\nonumber \\
&=& \frac{{{\gamma _{s{r_1},in}}}}{{{\gamma _{gap}}/2 + 1}} +
\frac{{{\gamma _{s{r_1},in}}}}{{{\gamma _{gap}}/3 + {\alpha _0}}} \nonumber \\
&=& {\gamma _{s{r_1},in}}\left( {\frac{{5/3{\gamma _{gap}} + 2(1 +
{\alpha _0})}}{{\left( {{\gamma _{gap}} + 2} \right)\left( {{\gamma
_{gap}}/3 + {\alpha _0}} \right)}}} \right).~~
\end{eqnarray}

For the optimal pairing, we have
\begin{eqnarray}
\rho _1^{opt}\left( {d_{(1)},{\gamma _{rd}},{\gamma
_{s{r_{(1)}},in}}} \right) &=& \frac{{\gamma _{gap}^{ - 1}{\gamma
_{s{r_1},in}}{\gamma _{s{r_1},in}}}}{{\left( {{\gamma
_{s{r_1},in}}d_{(2)}^{ - 1} + \gamma _{gap}^{ - 1}{\gamma
_{s{r_1},in}}} \right)}} \nonumber \\
&=& \frac{{{\gamma _{s{r_1},in}}}}{{\gamma _{gap}/3 + 1}};
\end{eqnarray}
\begin{eqnarray}
\rho _2^{opt}\left( {d_{(1)},{\gamma _{rd}},{\gamma
_{s{r_{(2)}},in}}} \right) &=& \frac{{\gamma _{gap}^{ - 1}{\gamma
_{s{r_1},in}}{\gamma _{s{r_2},in}}}}{{\left( {{\gamma
_{s{r_2},in}}d_{(1)}^{ - 1} + \gamma _{gap}^{ - 1}{\gamma
_{s{r_1},in}}} \right)}} \nonumber \\
&=& \frac{{{\gamma _{s{r_1},in}}}}{{\gamma _{gap}/2 + {\alpha _0}}};
\end{eqnarray}
\begin{eqnarray}
\rho^{opt} &=& \sum\limits_{k = 1}^2 {\rho _k^{opt}\left(
{d_{(k)}^{},{\gamma _{rd}},{\gamma _{s{r_{(k)}},in}}} \right)}  \nonumber \\
&=& \frac{{{\gamma _{s{r_1},in}}}}{{{\gamma _{gap}}/3 + 1}} +
\frac{{{\gamma _{s{r_1},in}}}}{{{\gamma _{gap}}/2 + {\alpha _0}}} \nonumber \\
&=& {\gamma _{s{r_1},in}}\left( {\frac{{5/3{\gamma _{gap}} + 2(1 +
{\alpha _0})}}{{\left( {{\gamma _{gap}}/3 + 1} \right)\left(
{{\gamma _{gap}} + 2{\alpha _0}} \right)}}} \right)
\end{eqnarray}

Similarly, for the conventional soft information relaying scheme,
${d_1} = {d_2} = 1$, and we get
\begin{eqnarray}
\rho _1^{SIR}\left( {1,{\gamma _{rd}},{\gamma _{s{r_1},in}}} \right)
&=& \frac{{\gamma _{gap}^{ - 1}{\gamma _{s{r_1},in}}{\gamma
_{s{r_1},in}}}}{{\left( {{\gamma _{s{r_1},in}} + \gamma _{gap}^{ -
1}{\gamma _{s{r_1},in}}} \right)}} \nonumber \\
&=& \frac{{{\gamma _{s{r_1},in}}}}{{\left( {\gamma _{gap} + 1}
\right)}};
\end{eqnarray}
\begin{eqnarray}
\rho _2^{SIR}\left( {1,{\gamma _{rd}},{\gamma _{s{r_2},in}}} \right)
&=& \frac{{\gamma _{gap}^{ - 1}{\gamma _{s{r_1},in}}{\gamma
_{s{r_2},in}}}}{{\left( {{\gamma _{s{r_2},in}} + \gamma _{gap}^{ -
1}{\gamma _{s{r_1},in}}} \right)}} \nonumber \\
&=& \frac{{{\gamma _{s{r_1},in}}}}{{\left( {\gamma _{gap} + {\alpha
_0}} \right)}};
\end{eqnarray}
\begin{eqnarray}
\rho^{SIR} &=& \sum\limits_{k = 1}^2 {\rho _k^{SIR}\left(
{d_k,{\gamma _{rd}},{\gamma _{s{r_k},in}}} \right)}  \nonumber \\
&=& \frac{{{\gamma _{s{r_1},in}}}}{{\left( {\gamma _{gap}^{} + 1}
\right)}} + \frac{{{\gamma _{s{r_1},in}}}}{{\left( {\gamma _{gap}^{}
+ {\alpha _0}} \right)}} \nonumber \\
&=& {\gamma _{s{r_1},in}}\left( {\frac{{2{\gamma _{gap}} + (1 +
{\alpha _0})}}{{\left( {{\gamma _{gap}} + 1} \right)\left( {{\gamma
_{gap}} + {\alpha _0}} \right)}}} \right).
\end{eqnarray}

Then we have
\begin{eqnarray}
\rho^{opt} - \rho^{un - order} = ~~~~~~~~~~~~~~~~~~~~~~~~~~~~~~~~~~~~~~~~~~~~ \nonumber\\
\frac{{\gamma _{gap}\left( {{\alpha _0} - 1} \right)\left( {5{\gamma
_{gap}} + 6(1 + {\alpha _0})} \right)}}{{\left( {{\gamma _{gap}} +
3} \right)\left( {{\gamma _{gap}} + 2{\alpha _0}} \right)\left(
{{\gamma _{gap}} + 2} \right)\left( {{\gamma _{gap}} + 3{\alpha _0}}
\right)}}{\gamma _{s{r_1},in}} > 0, \nonumber
\end{eqnarray}
\begin{eqnarray}
\rho^{opt} - \rho^{SIR} =
~~~~~~~~~~~~~~~~~~~~~~~~~~~~~~~~~~~~~~~~~~~~~~~~ \nonumber \\
\frac{{\gamma _{gap}^{}\left( {3\gamma _{gap}^2 + (4 + 6{\alpha
_0}){\gamma _{gap}} + (4\alpha _0^2 + 3)} \right)}}{{\left( {{\gamma
_{gap}} + 3} \right)\left( {{\gamma _{gap}} + 2{\alpha _0}}
\right)\left( {{\gamma _{gap}} + 1} \right)\left( {{\gamma _{gap}} +
{\alpha _0}} \right)}}{\gamma _{s{r_1},in}} > 0. \nonumber
\end{eqnarray}

$\rho^{opt} - \rho^{un - order}$ and $\rho^{opt} - \rho^{SIR}$ are
the coding gains of the DISC with the optimal code pairing over the
DISC with an un-ordered pairing and the conventional SIR scheme,
respectively. From the above two equations, we can see that the DISC
with the optimal pairing outperforms the DISC with an un-ordered
pairing and the conventional SIR scheme at high SNR.

\begin{figure}
\hspace{-1.5cm}
\includegraphics[width=0.60\textwidth]{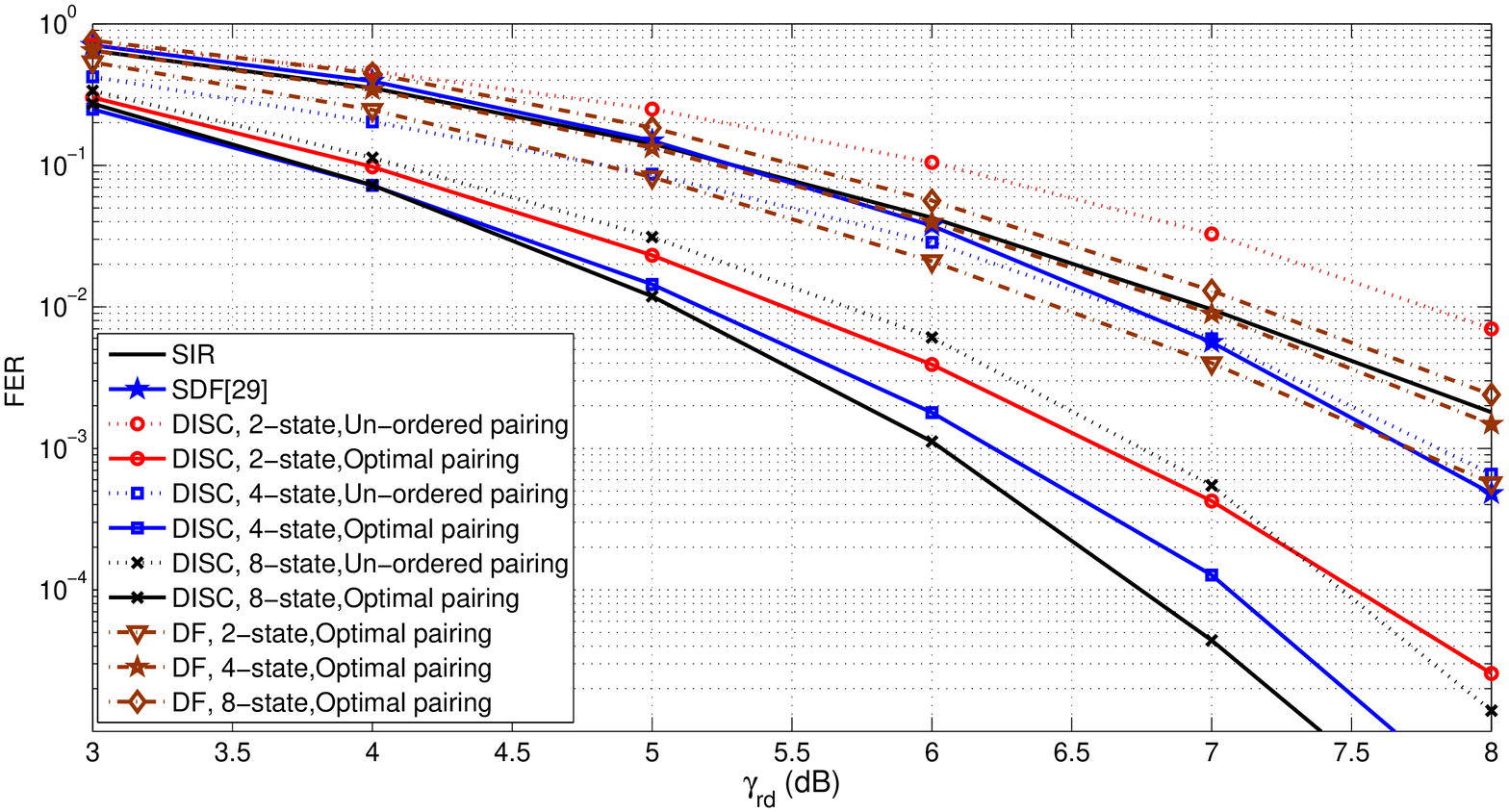}
\caption{FER for 2 relays in AWGN Channels, ${\bar \gamma _{sr}} =
{\bar \gamma _{s{r_1}}} = {\bar \gamma _{s{r_2}}} - 3dB$, ${\bar
\gamma _{sr}} = {\bar \gamma _{rd}}$.}
\end{figure}

\begin{figure}
\hspace{-0.5cm}
\includegraphics[width=0.6\textwidth]{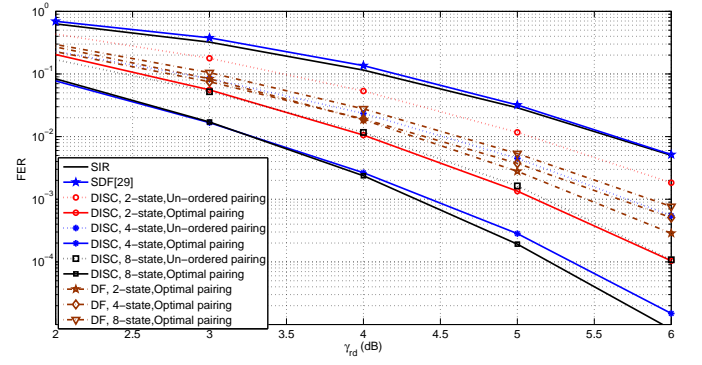}
\caption{FER for 2 relays in AWGN channels, ${\bar \gamma _{sr}} =
{\bar \gamma _{s{r_1}}} = {\bar \gamma _{s{r_2}}} - 3dB$, ${\bar
\gamma _{sr}} = {\bar \gamma _{rd}} + 3dB$.}
\end{figure}

\section{Simulation Results and Discussions}

In this section, we present the simulation results. All simulations
are performed for the BPSK modulation and a frame size of 130
symbols over AWGN and quasi-static fading channels. We assume that
all relays use the code with the same number of states. Thus if
there are K relays, the convolutional code of rate 1/K with K
constituent codes can be used in K relays, and each relay uses one
constituent code. For example, for the relay network with 2, 3 and 4
relays, the convolutional codes listed in Table 1 can be used as
constituent codes at the relays. All schemes are compared under the
same power constraint.

Figs. 5-6 compare the frame error rate (FER) performance of the
proposed DISC with optimum and un-ordered code pairing, soft
information relaying (SIR) [6], soft decode and forward (SDF) [29]
as well as the conventional detect and forward (DF) with relay
re-encoding (DF) for various numbers of states over AWGN channels
for ${\bar \gamma _{sr}} = {\bar \gamma _{rd}}$, ${\bar \gamma
_{sr}} = {\bar \gamma _{rd}}$+3dB, respectively. We set ${\bar
\gamma _{s{r_1}}} = {\bar \gamma _{sr}}$ and ${\bar \gamma
_{s{r_2}}} = {\bar \gamma _{sr}} + 3dB$. Here the optimal and
un-ordered assignment is the same as the assignment given in Example
2. We assume in the SIR scheme that the overall destination noise
follows the Gaussian distribution and the maximum ratio combining
(MRC) is used to combine received signals from all relays.

We can first note from these figures that the performance of the DF
with relay encoding gets worse as the number of relay encoder states
increases. Such performance degradation is due to the error
propagations in the DF scheme. In the DF scheme, when detection
errors occur at the relays, the re-encoding process causes errors to
propagate into subsequent symbols. The longer the encoder memory,
the larger the number of subsequent symbols affected by the
detection errors. Therefore, the error rate of the DF will increase
with the number of states. For example, Fig. 5 shows that the FER
performance of the DF for the 4-state and 8-state codes is worse by
0.5dB and 0.7dB, respectively, than for the 2-state code at the FER
of $10^{-3}$. However, the SISO encoder in the proposed DISC scheme
can effectively mitigate the error propagation in the relay encoding
process and at the same time provide a significant distributed
coding gain. As a result, the DISC provides significant coding gains
compared to the SIR without relay encoding and SDF scheme [29]. The
gain increases as the number of states increases at high SNR. For
example, as shown in Fig. 6, the DISCs with 2, 4 and 8-state are
superior to the SIR and SDF scheme by about 1.7dB, 2.4dB and 2.5dB
respectively. This result is consistent with the analysis in Section
3, showing that the DISC performance improves as the number of
states at the relay encoder increases. Furthermore, we can also see
from the figures that the DISC with optimal code pairing also brings
significant gains compared to the DISC with the un-ordered pairing.
For example, as shown in Fig. 5, the 4-state code with the optimal
code pairing is superior to that with the un-ordered pairing by 2dB
at the FER of $10^{-3}$. This validates the effectiveness of the
proposed design criteria. Since the performance of the DISC with the
un-ordered code pairing is exactly the same as the SISO encoding
scheme in [8], the proposed scheme significantly outperforms the
SISO encoding scheme in [8] and the gain comes from the optimal code
paring.

\begin{figure}
\hspace{-1.5cm}
\includegraphics[width=0.6\textwidth]{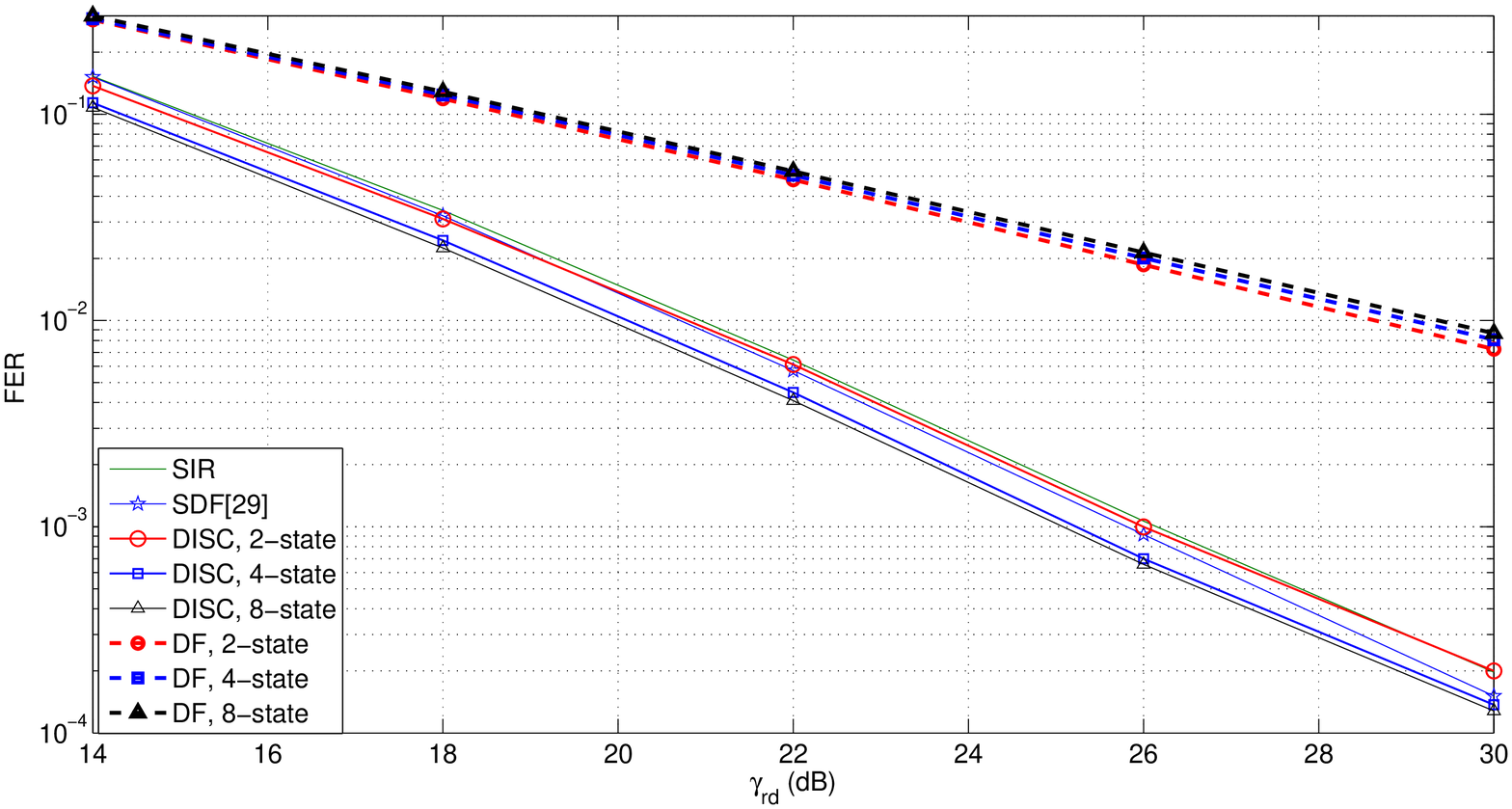}
\caption{FER for 2 relays in fading channels, ${\bar \gamma _{sr}} =
{\bar \gamma _{s{r_1}}} = {\bar \gamma _{s{r_2}}}$, ${\bar \gamma
_{sr}} = {\bar \gamma _{rd}}$.}
\end{figure}

\begin{figure}
\hspace{-1.5cm}
\includegraphics[width=0.6\textwidth]{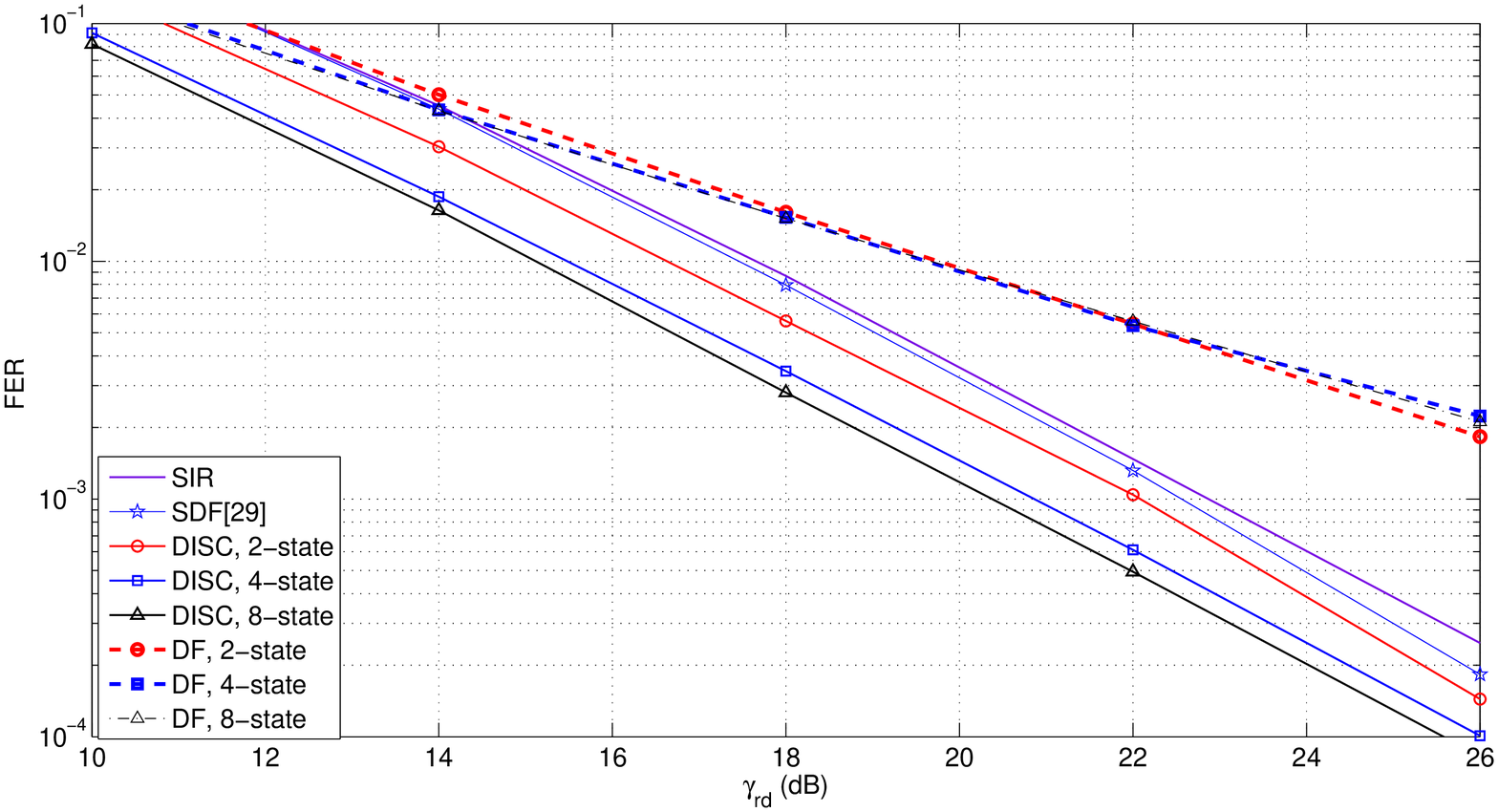}
\caption{FER for 2 relays in fading channels, ${\bar \gamma _{sr}} =
{\bar \gamma _{s{r_1}}} = {\bar \gamma _{s{r_2}}},{\bar \gamma
_{sr}} = {\bar \gamma _{rd}} + 10dB$.}
\end{figure}



Figs. 7-8 compare the performance over fading channels for a network
with 2 relays for  ${\bar \gamma _{sr}} = {\bar \gamma _{rd}}$,
${\bar \gamma _{sr}} = {\bar \gamma _{rd}}$ +10dB, respectively.  We
set ${\bar \gamma _{s{r_1}}} = {\bar \gamma _{s{r_2}}} = {\bar
\gamma _{sr}}$. It can also be observed from the figures that the
DISC and SIR can achieve the full diversity order of 2, but the
conventional DF can only achieve the diversity order of 1 due to
error propagation. These two figures also compare the performance of
DISC with SIR without relay re-encoding. It can be seen that the
DISC substantially outperforms SIR. For example, the DISC with 2, 4
and 8 states can bring about 0.1dB, 0.8dB and 1dB gains,
respectively, relative to the SIR scheme, for ${\bar \gamma _{sr}} =
{\bar \gamma _{rd}}$ and the gains are increased to 1dB, 2dB and
2.5dB, respectively, for ${\bar \gamma _{sr}} = {\bar \gamma
_{rd}}$+10dB. That is, the coding gain brought by the DISC increases
when the source-relay link quality is improved. We can also observe
that the coding gain increases as the number of relay encoder states
increases.

\begin{figure}
\hspace{-0.5cm}
\includegraphics[width=0.6\textwidth]{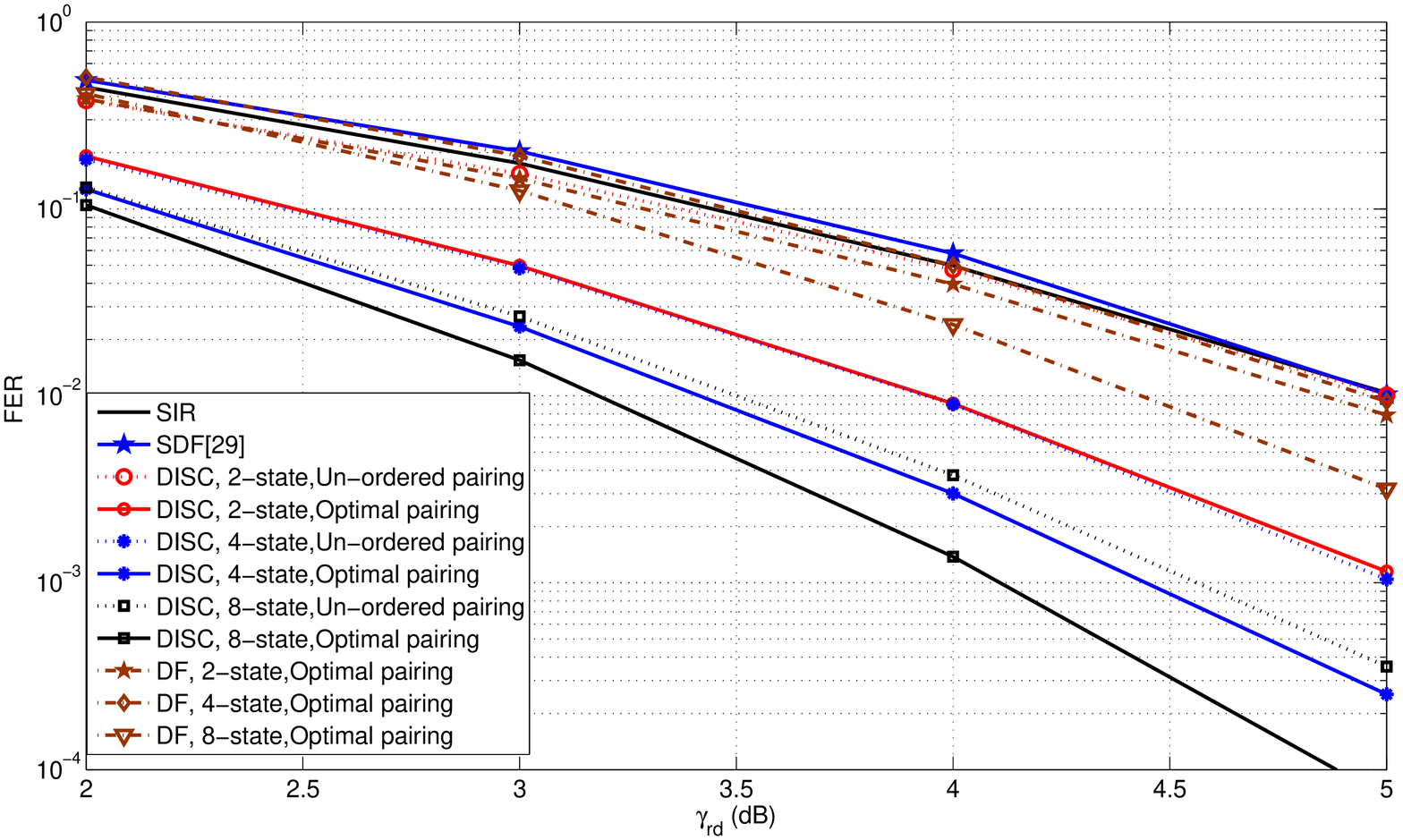}
\caption{FER for 3 relays in AWGN Channels, ${\bar \gamma _{sr}} =
{\bar \gamma _{s{r_1}}} = {\bar \gamma _{s{r_2}}} - 2dB = {\bar
\gamma _{s{r_3}}} - 4dB$, ${\bar \gamma _{sr}} = {\bar \gamma
_{rd}}$.}
\end{figure}

\begin{figure}
\hspace{-0.5cm}
\includegraphics[width=0.6\textwidth]{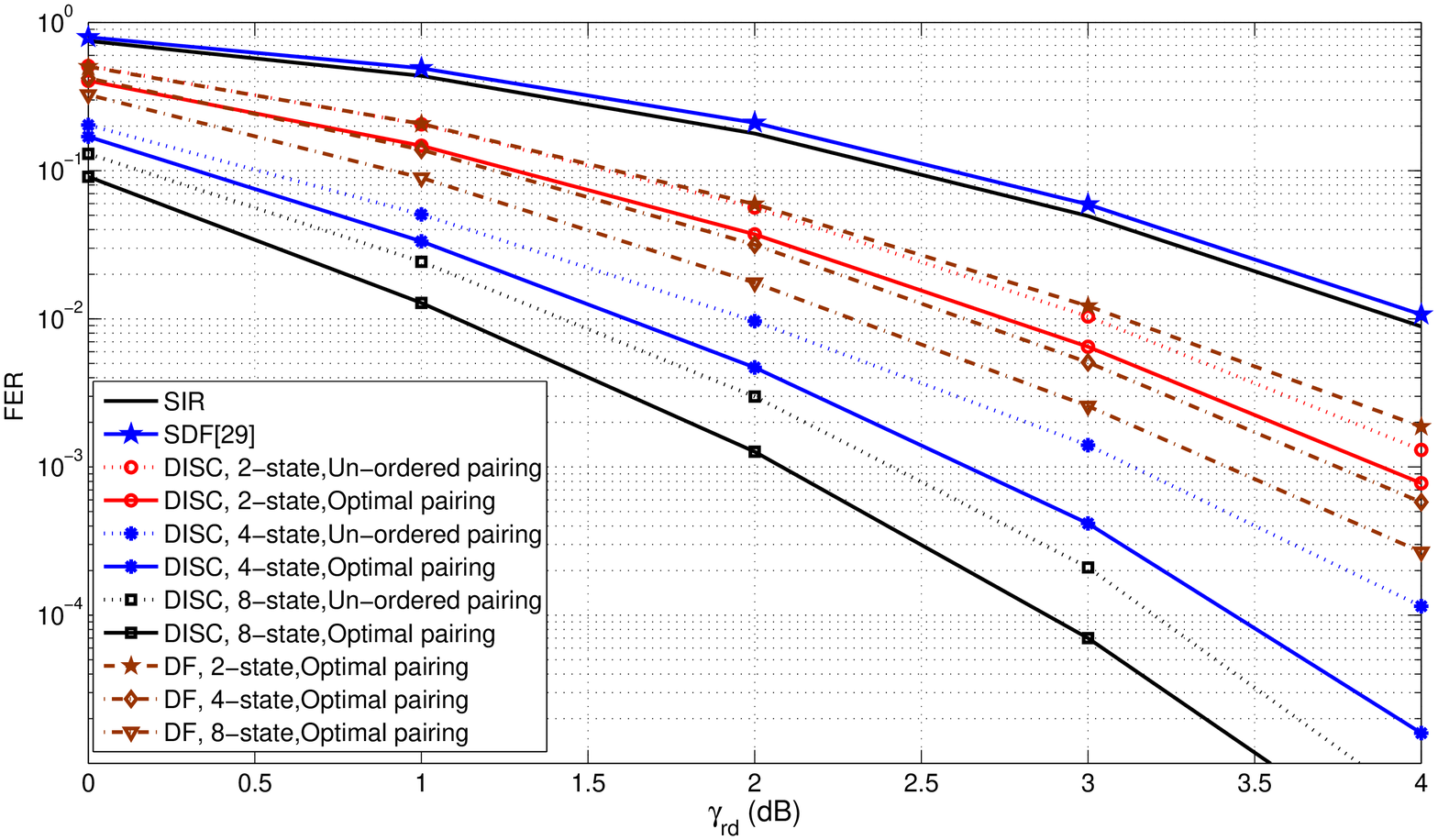}
\caption{FER for 3 relays in AWGN Channels, ${\bar \gamma _{sr}} =
{\bar \gamma _{sr,1}} = {\bar \gamma _{s{r_2}}} - 2dB = {\bar \gamma
_{s{r_3}}} - 4dB$, ${\bar \gamma _{sr}} = {\bar \gamma _{rd}} +
3dB$.}
\end{figure}


Figs. 9-12 show the results for three relays over AWGN and fading
channels, respectively. From these figures we can observe similar
trends as for the case with two relays. That is, the DISC can bring
significant gains over the SIR, SDF and DF with re-encoding on both
AWGN and fading channels and the gains increase as the number of
state increases. For the fading channels, both DISC and SIR schemes
can achieve the full diversity order of 3 while the DF can only
achieve the diversity order of 1. Also pairing can bring system
considerable gains over AWGN channels. Furthermore, the coding gain
of DISC over SIR slightly increases as the number of relay increases
from 2 to 3.

From the above results, we can see that the relay encoding in the
conventional DF schemes cause serious error propagation and thus
does not provide any coding advantages. By contrary, the proposed
DISC can effectively mitigate the error propagation in the relay
encoding and provide significant distributed coding gains, thus
substantially outperforming the soft information relaying (SIR) and
conventional DF schemes.

\begin{figure}
\hspace{-1.5cm}
\includegraphics[width=0.6\textwidth]{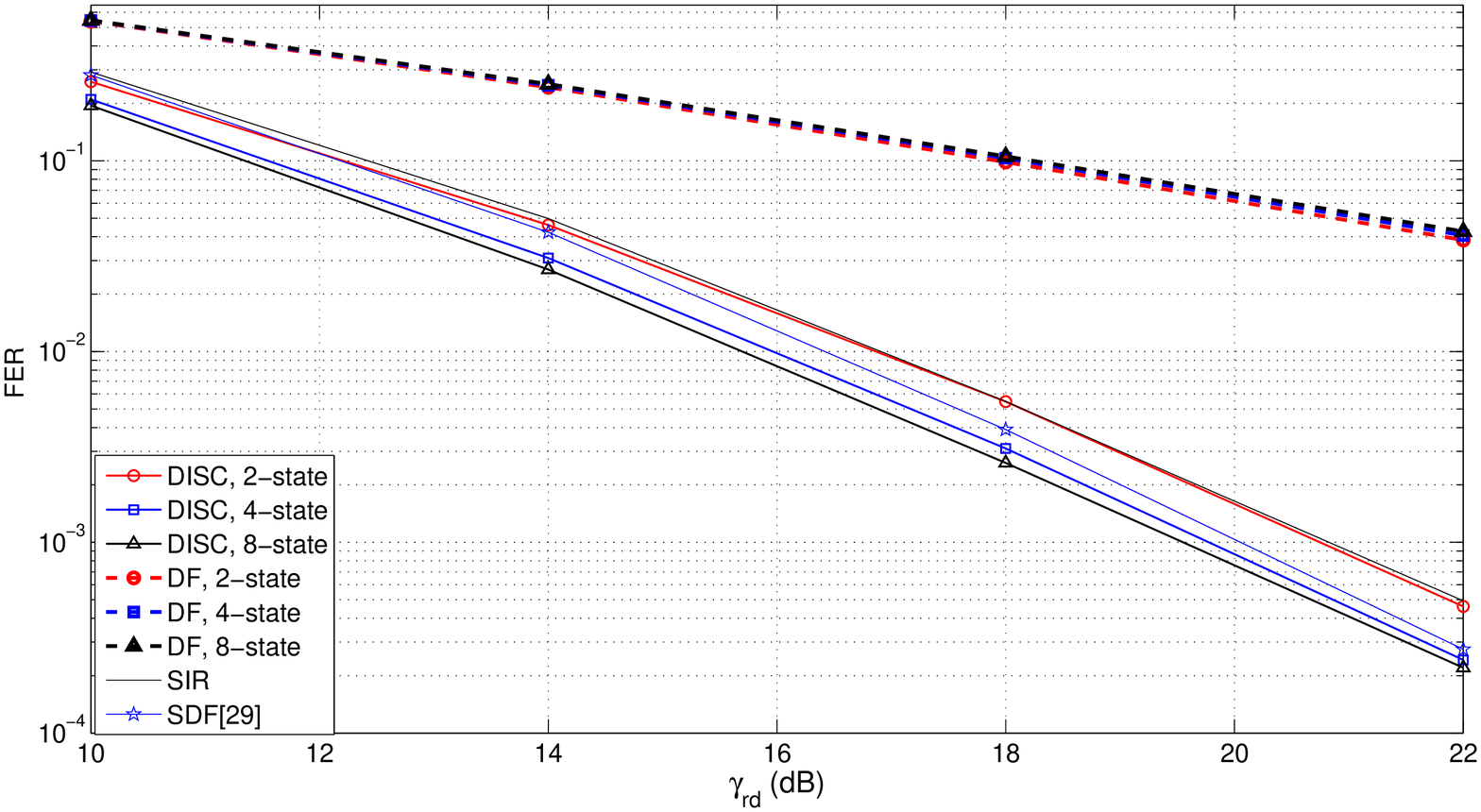}
\caption{FER for 3 relays in fading channels, ${\bar \gamma _{sr}} =
{\bar \gamma _{sr,1}} = {\bar \gamma _{s{r_2}}} = {\bar \gamma
_{s{r_3}}},{\bar \gamma _{sr}} = {\bar \gamma _{rd}}$.}
\end{figure}

\begin{figure}
\hspace{-1.5cm}
\includegraphics[width=0.6\textwidth]{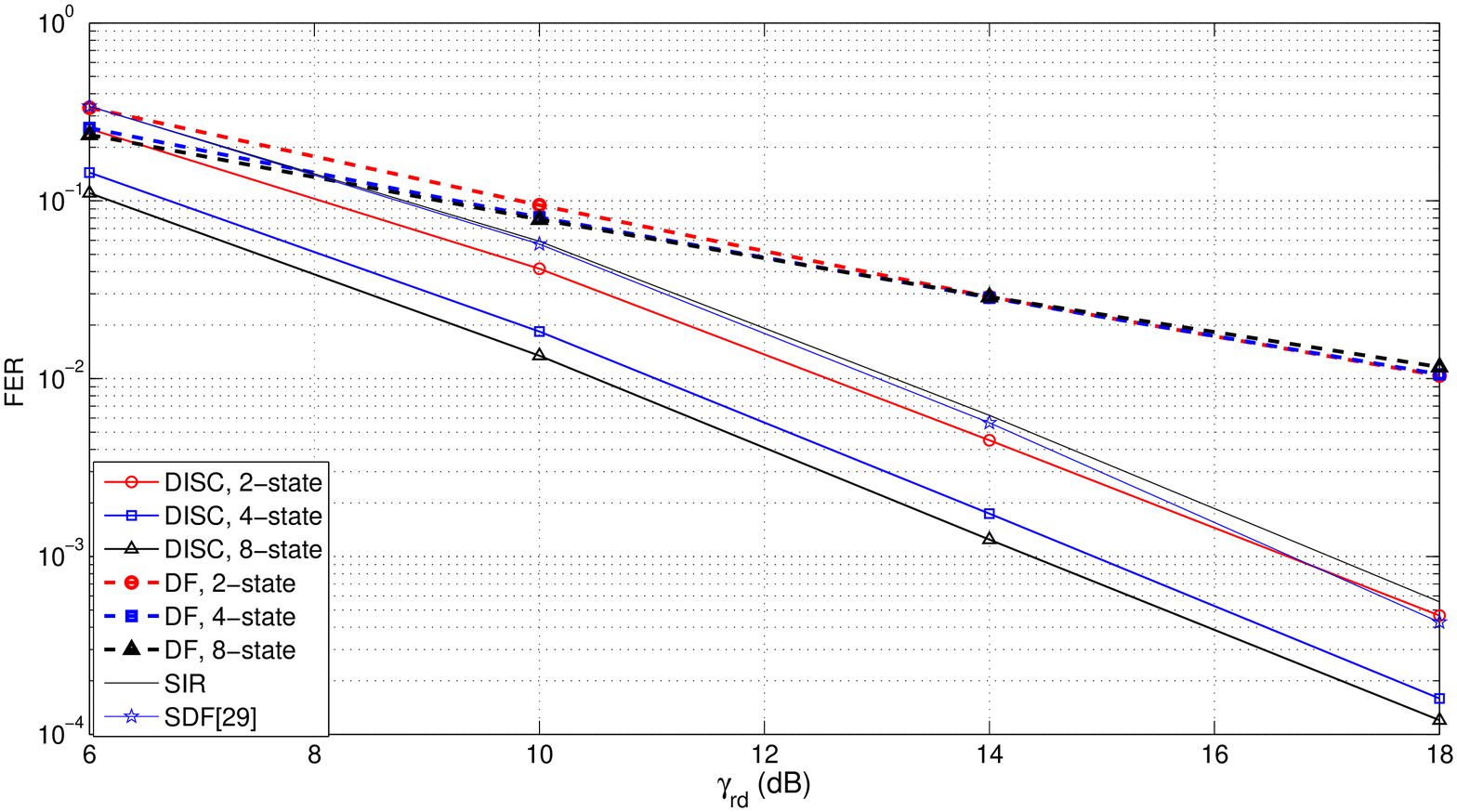}
\caption{FER for 3 relays in fading channels, ${\bar \gamma _{sr}} =
{\bar \gamma _{sr,1}} = {\bar \gamma _{s{r_2}}} = {\bar \gamma
_{s{r_3}}},{\bar \gamma _{sr}} = {\bar \gamma _{rd}} + 10dB$.}
\end{figure}

\section{Conclusions}

In this paper we proposed a new distributed soft coding (DISC)
scheme based on a soft input soft output encoder at the relay and
the optimal design criteria were proposed. The proposed scheme
performs encoding of the noisy source symbol estimates in
error-prone relay channels. It can effectively circumvent error
propagation in the relay encoding process and at the same time
provide significant distributed coding gains. It provides great
performance improvement compared to the soft information relaying
(SIR) and conventional DF schemes. At high SNR the gain further
increases as the number of state increases.

In the analysis and simulations of DISC in this paper we
approximated the distribution of the overall destination noise as
the Gaussian distribution. As shown in [7, 34], the actual
distribution does not follow the exact Gaussian distribution.
Further improvement can be expected when the accurate distribution
of the overall destination noise is derived, but unfortunately it
has been a quite challenging topic so far.

\section{Acknowledgement}

We would like to all the reviewers for their valuable comments and
suggestions for improving the quality of the paper. We also would
like to thank Dr. Ingmar Land for the thoughtful discussions about
the error modeling in DF scheme.

\section{Appendix}

\subsection{Proof of Theorem 1}

To prove this theorem, let us first prove the following Lemma.

\textbf{Lemma 1}: Let $c(i)$ be the $i$-th output symbol of a binary
encoder generated by $\mathbf{G}$ for source binary sequence
$\bf{b}=\left({b(1),\cdots,b(N)}\right)$. It is given by
\begin{eqnarray}
c(i) = {{\bf{G}}_i}{{\bf{b}}^T} = \sum\limits_{n = 1}^N
{G_{in}b(n)},
\end{eqnarray}
where the summation is over GF(2). Let ${l_b}(n) = \ln \left( {\Pr
(b(n) = 0)/\Pr (b(n) = 1} \right)$ represent the log likelihood
ratio (LLR) of $b(n)$. Then the LLR of $c(i)$, denoted by
${l_c}\left( i \right)$, can be calculated as follows
\begin{eqnarray}
{l_c}\left( i \right) = \ln \frac{{\Pr (c(i) = 0)}}{{\Pr (c(i) =
1)}} = \ln \frac{{1 + \prod\nolimits_{n \in {U_i}}{\tanh \left(
{{l_b}(n)/2} \right)} }}{{1 - \prod\nolimits_{n \in {U_i}}{\tanh
\left( {{l_b}(n)/2} \right)} }}, \label{eq37a}
\end{eqnarray}
where $\tanh (x) = \frac{{{e^{2x}} - 1}}{{{e^{2x}} + 1}}$.

The Lemma 1 can be directly proved by using the L-algebra [19]
\begin{eqnarray}
l\left( {b({n_1}) \oplus b({n_2}) \cdots  \oplus b({n_k})} \right)
~~~~~~~\nonumber \\
= \ln L\left( {b({n_1}) \oplus b({n_2}) \cdots  \oplus b({n_k})}
\right) \nonumber \\ = \ln \frac{{1 + \prod\nolimits_{q = 1}^k
{\tanh ({l_b}({n_q})/2)} }}{{1 - \prod\nolimits_{q = 1}^k {\tanh
({l_b}({n_q})/2)} }}, ~~~ ~\label{eq14}
\end{eqnarray}
where $L(x)$ and $l(x)$ denote the likelihood ratio and LLR of $x$,
respectively.

From Lemma 1, we can easily prove the Theorem 1. Let ${\tilde
x_b}(k)$ and  ${l_b}\left( k \right)$ represent the SBE and LLR of
source bit $b(k)$. Then they have the following relationship,
\begin{eqnarray}
{l_b}\left( k \right) = \ln \frac{{1 + {{\tilde x}_b}(k)}}{{1 -
{{\tilde x}_b}(k)}}, \label{eq36}
\end{eqnarray}
\begin{eqnarray}
{\tilde x_b}(k) = \tanh \left( {{l_b}(k)/2} \right) =\frac{{\exp
\left( {{l_b}(k)} \right) - 1}}{{\exp \left( {{l_b}(k)} \right) +
1}}.  \label{eq37}
\end{eqnarray}

By substituting Eq. (\ref{eq37}) into (\ref{eq37a}), we have
\begin{eqnarray}
{l_c}\left( i \right) = \ln \frac{{1 + \prod\nolimits_{j \in {U_i}}
{{{\tilde x}_b}(j)} }}{{1 - \prod\nolimits_{j \in {U_i}} {{{\tilde
x}_b}(j)} }}. \label{eq38}
\end{eqnarray}

Then by using the relationship of the SBE and LLR in Eq.
(\ref{eq37}), the SBE of $c(k)$, denoted by ${\tilde x_c}(k)$, can
be calculated as
\begin{eqnarray}
{\tilde x_c}(k)=\prod\nolimits_{j \in {U_k}}{{{\tilde x}_b}(j)}  =
\exp \left( {{{\bf{G}}_k}{{\left( {{\bf{ln}}{{{\bf{\tilde x}}}_b}}
\right)}^T}} \right). \label{eq39}
\end{eqnarray}

This proves Theorem 1.

\subsection{Proof of Theorem 3}

To prove the theorem, we need to use the following Lemma.

\textbf{Lemma 2}: Given the positive real numbers ${v_1} > {v_2} >
0$ and ${\theta _1} > {\theta _2} > 0$, the following relationship
always holds,
\begin{eqnarray}
\frac{1}{{{v_1} + {\theta _1}}} + \frac{1}{{{v_2} + {\theta _2}}} >
\frac{1}{{{v_1} + {\theta _2}}} + \frac{1}{{{v_2} + {\theta _1}}}.
\label{eq41}
\end{eqnarray}

The proof of the above result is straightforward and we omit it
here. Now let us use the Lemma to prove Theorem 3.

Since ${\gamma _{r_k{d}}} = {\gamma _{rd}}$ for all $k=1,\ldots,K$,
we can rewrite ${\rho _k}\left( {d_k,{\gamma _{rd}},{\gamma
_{s{r_k},in}}} \right)$ as follows
\begin{eqnarray}
{\rho _k}\left( {d_k,{\gamma _{rd}},{\gamma _{s{r_k},in}}} \right) =
\frac{{{\gamma _{rd}}{\gamma _{s{r_k},in}}}}{{{\gamma
_{s{r_k},in}}d_k^{ - 1} + {\gamma _{rd}}}} \nonumber \\ =
\frac{1}{{{{\left( {d_k{\gamma _{rd}}} \right)}^{ - 1}} + {{\left(
{{\gamma _{sr_k,in}}} \right)}^{ - 1}}}} = \frac{1}{{\left( {{v_k} +
{\theta _k}} \right)}}, \label{eq42}
\end{eqnarray}
where ${v_k} = {\left( {d_k{\gamma _{rd}}} \right)^{ - 1}}$ and
${\theta _k} = {\left( {{\gamma _{s{r_k},in}}} \right)^{ - 1}}$.

Let ${v_{(1)}}$, ${v_{(2)}}$, \ldots,${v_{(K)}}$ and
${\theta_{(1)}}$, ${\theta_{(2)}}$, \ldots,${\theta_{(K)}}$
represent the re-ordered values of ${v_{1}}$, ${v_{2}}$,
\ldots,${v_{K}}$ and ${\theta_{1}}$, ${\theta_{2}}$,
\ldots,${\theta_{K}}$. Then from Eqs. (\ref{eq32}) and (\ref{eq33}),
we have
\begin{eqnarray}
{v_{(1)}} \le {v_{(2)}} \cdots  \le {v_{(K)}}~and~{\theta _{(1)}}
\le {\theta _{(2)}} \cdots  \le {\theta _{(K)}}. \label{eq43}
\end{eqnarray}

Now let us determine how to distribute $\{ {v_{(1)}}, \cdots
,{v_{(K)}}\}$ and  $\{{\theta _{(1)}}, \cdots ,{\theta _{(K)}}\}$ to
form $K$ pairs $\left( {{v_k},{\theta _k}} \right)$, $k=1,\ldots,K$,
where any ${v_{(i)}}$ or ${\theta _{(j)}}$ can only be assigned to
one and only one pair, so as to maximize $\sum\limits_{k = 1}^K
{{\rho _k}\left( {d_k^{},{\gamma _{rd}},{\gamma _{sr_k,in}}}
\right)}$. As can be seen from Eq. (\ref{eq42}), this is equivalent
to maximizing
\begin{eqnarray}
{\Sigma _K} = \sum\limits_{k = 1}^K {\frac{1}{{{v_k} + {\theta
_k}}}}  = \sum\limits_{k = 1}^K {f\left( {{v_k},{\theta _k}}
\right)}, \label{eq44}
\end{eqnarray}
where $f\left( {{v_k},{\theta _k}} \right) = {({v_k} + {\theta
_k})^{ - 1}}$.

We assume that the optimal $K$ pairs of   for $k=1,\ldots,K$, are
$\left( {{v_{(1)}},{\theta _{{j_1}}}} \right)$, $\left(
{{v_{(2)}},{\theta _{{j_2}}}} \right)$,$\ldots$, $\left(
{{v_{(K)}},{\theta _{{j_K}}}} \right)$. Now we prove this theorem by
contradiction. Assume that ${\theta _{{j_1}}}, {\theta _{{j_2}}},
\ldots, {\theta _{{j_K}}}$ do not exactly follow the relationship of
${\theta _{{j_1}}} \le {\theta _{{j_2}}} \cdots  \le {\theta
_{{j_K}}}$. Then there must exist at least two integers $p, q$, such
that ${\theta _{{j_p}}} > {\theta _{{j_q}}}$ for $p<q$. Since
${v_{(q)}} \ge {v_{(p)}}$ and ${\theta _{{j_p}}} > {\theta
_{{j_q}}}$, then by using the Lemma 1, we have
\begin{eqnarray}
f\left( {{v_{(q)}},{q_{{j_p}}}} \right) + f\left( {{v_{(p)}},{\theta
_{{j_q}}}} \right) > f\left( {{v_{(q)}},{\theta _{{j_q}}}} \right) +
f\left( {{v_{(p)}},{\theta _{{j_p}}}} \right). \label{eq45}
\end{eqnarray}

This means that when we switch  ${\theta _{{j_q}}}$ and ${\theta
_{{j_p}}}$ in the two pairs $\left( {{v_{(p)}},{\theta _{{j_p}}}}
\right)$  and $\left( {{v_{(q)}},{\theta _{{j_q}}}} \right)$,
leading to a new two pairs $\left( {{v_{(p)}},{\theta _{{j_q}}}}
\right)$ and $\left( {{v_{(q)}},{\theta _{{j_p}}}} \right)$, while
keeping other pairs unchanged, the resulted ${\Sigma _K}$ will be
larger. This contradicts that  $\left( {{{\bf{v}}_{(1)}},{\theta
_{{j_1}}}} \right), \cdots,  \left( {{v_{(K)}},{\theta _{{j_K}}}}
\right)$ are the optimal pairs achieving the maximum ${\Sigma _K}$.
Therefore, the optimal pairs are  $\left( {{v_{(1)}},{\theta
_{(1)}}} \right),  \left( {{v_{(2)}},{\theta _{(2)}}}
\right),\cdots, \left( {{v_{(K)}},{\theta _{(K)}}} \right)$. This
proves Theorem 3.

\bibliographystyle{IEEE}

\end{document}